\documentclass[acmsmall]{acmart}
\AtBeginDocument{%
  \providecommand\BibTeX{{%
    \normalfont B\kern-0.5em{\scshape i\kern-0.25em b}\kern-0.8em\TeX}}}

\setcopyright{acmcopyright}
\copyrightyear{2023}
\acmYear{2023}
\acmDOI{XXXXXXX.XXXXXXX}

\newcommand{\s}{\\}
\newcommand{\bs}{\boldsymbol}

\newcommand{\mbf}{\mathbf}
\newcommand{\jac}{J_{\mbf u} \,}

\newcommand{\evs}[1]{\textcolor{black}{#1}}
\newcommand{\arc}[1]{\textcolor{black}{#1}}
\DeclareMathOperator{\range}{im}
\usepackage{physics}
\usepackage{amsmath}
\usepackage{mathtools}
\newcommand{\innerp}[2]{\left \langle #1, \, #2 \right \rangle}
\usepackage{algorithm}
\usepackage{algorithmic}

\acmJournal{TOMS}
\acmVolume{**}
\acmNumber{*}
\acmArticle{***}
\acmMonth{*}

\begin{document}

\setcopyright{acmlicensed}
\acmJournal{TOMS}
\acmYear{2023} \acmVolume{1} \acmNumber{1} \acmArticle{1} \acmMonth{1} \acmPrice{}\acmDOI{10.1145/3625560}

\title{Computation of Turing bifurcation normal form for $n$-component reaction-diffusion systems.}

\author{Edgardo Villar-Sep\'ulveda}
\affiliation{%
  \institution{University of Bristol}
  \streetaddress{Department of Engineering Mathematics, University of Bristol}
  \city{Bristol BS8 1TW}
  \country{United Kingdom}}
\email{edgardo.villar-sepulveda@bristol.ac.uk}

\author{Alan Champneys}
\affiliation{%
  \institution{University of Bristol}
  \city{Bristol}
  \country{United Kingdom}
}

\renewcommand{\shortauthors}{Villar-Sep\'ulveda \& Champneys}

\begin{abstract}
    General expressions are derived for the amplitude equation valid at a Turing bifurcation of a system of reaction-diffusion equations in one spatial dimension, with an arbitrary number of components. The normal form is computed up to fifth order, which enables the detection and analysis of codimension-two points where the criticality of the bifurcation changes. The expressions are implemented within a Python package, in which the user needs to specify only expressions for the reaction kinetics and the values of diffusion constants. The code is augmented with a Mathematica routine to compute curves of Turing bifurcations in a parameter plane and automatically detect codimension-two points. The software is illustrated with examples that show the versatility of the method including a case with cross-diffusion, a higher-order scalar equation, and a four-component system.
\end{abstract}



\maketitle

\section{Introduction}
    Turing instability is a fundamental mechanism for the formation of patterns in spatially extended reaction-diffusion systems arising across life and physical sciences, see \cite{Murray2, Meron} and references therein. The theory and application of such instabilities remain active areas of research, see e.g.~\cite{Krause} for a recent overview.
    
    Recently, several authors have shown the connection between Turing instability and the birth of localised states in scalar pattern-forming models such as the Swift-Hohenberg equation, see \cite{Knobloch_review, MeronIssue}. On a long domain, Turing bifurcation can be considered to be a pitchfork-like bifurcation of the amplitude of sinusoidal-like patterns corresponding to a critical wave number $k$ of the dispersion relation. 
    
    A distinction can be drawn between cases giving rise to small-amplitude stable periodic patterns and those leading to localised structures according to whether the corresponding bifurcation is super-\ or sub-critical. 
    
    In a companion to the present paper \cite{Villar1}, we developed necessary and sufficient conditions for an $n$-component reaction-diffusion system to undergo either a Turing bifurcation (corresponding to a critical wave number with a zero temporal eigenvalue) or a wave bifurcation (corresponding to a pure imaginary pair of temporal eigenvalues). That work though does not enable one to determine the criticality of the bifurcation. The purpose of this paper then, is to determine that for a Turing bifurcation occurring in a quite general reaction-diffusion system limited to one spatial dimension.
    
    In finite-dimensional dynamical systems, several different styles of normal forms enable the analysis of small-amplitude solutions emanating from a local bifurcation, see \cite{Murdock,kuznetsov,GuHo:83}. In infinite-dimensional systems, whose dynamics are governed by partial differential equations (PDEs), there are several other approaches for describing small-amplitude bifurcating solutions, depending on the function space with which we are working. In turn, the function space typically depends on the domain, boundary conditions, and other constraints. 
    
    Here though, we take a different point of view, that of the so-called spatial dynamics \cite{Haragus} for PDEs in one spatial dimension, where one considers the domain to be the real line, requiring solutions only to be bounded as $x \to \pm \infty$. Within spatial dynamics, the fundamental pattern-forming bifurcation is known as a Hamiltonian-Hopf, or reversible 1:1 resonance bifurcation, which can be shown \cite{Victor} to be the accumulation point of Turing bifurcations that would occur on a finite domain subject to, \evs{for example, Neumann or periodic} boundary conditions.
        
    Within this context, many different methods can be used to determine the appropriate normal forms, for example, singularity and group theory \cite{golubitsky}, or the method of multiple scales \evs{\cite{Coullet}. To compute the coefficients of the normal form, for a particular example, traditionally one might first perform centre manifold reduction, see e.g. \cite{GuHo:83}. However, more recently, authors have shown that applying the method of multiple scales directly, without centre manifold reduction\arc{,} is equivalent, see e.g.~\cite{KuznetsovNF}.} Here, we shall adopt the latter method, as described by Elphick {\em et al} \cite{tirapegui}, adapted for 2-component reaction-diffusion systems in \cite{fahad}, and for the Swift-Hohenberg fourth-order scalar equation in \cite{knobloch}. \evs{The method is also similar to the approach used by Ipsen {\em et al} \cite{Ipsen2} for computation of normal forms at bifurcations}.
    
    \evs{In this paper, we restrict attention to PDE systems that can be written as reaction-diffusion equations on the real line. In that case, the normal form of the Turing bifurcation becomes that of a pitchfork bifurcation of revolution (see Eq.~\eqref{eq:NFeqn} below), which can be super or sub-critical depending on the sign of the third-order \evs{coefficient} ($C_3$ in the notation used below).}  In the case that the bifurcation is sub-critical (\evs{$C_3 > 0$}) localised patterns occur only under an appropriate sign condition of the fifth-order term \cite{IoPe}. Under this sign condition ($C_5 < 0$ in our notation) the vanishing of the third-order coefficient can lead to a codimension-two bifurcation that seems to act as an organising centre, giving birth to localised patterns in reaction-diffusion systems \cite{MeronIssue,fahad}.
    
    \evs{If one is interested in the full dynamics in a neighbourhood of the codimension-two point where $C_3 = 0$, then further terms arise, in what we call an {\em extended} normal form, see \cite{ponedel2017front} and Sec.~\ref{sec:3.6} below. The full unfolding is also affected by terms that go beyond all orders of the normal form approximation, see e.g.~\cite{Kozyreff,Dean,deWitt}. Indeed only under certain conditions (in addition to $C_3\neq 0$) can the dynamics of the
    normal form be rigorously shown to be topologically conjugate to the dynamics of the full PDE system, on an appropriate centre manifold; for example, if the PDE is posed on the real line or a finite domain with periodic boundary conditions, see e.g.~\cite{Haragus}. Such considerations of rigorous equivalence go beyond the scope of this paper.}
        
    \evs{There has been considerably less work on deriving normal forms for Turing bifurcations in systems of reaction-diffusion equations with more than two components. Such systems arise for example} in the context of chemical physics \cite{Purwins} and cell biology \cite{Jilkine}. \evs{Ipsen {\em et al} \cite{Ipsen1,Ipsen3} provided general expressions for cubic coefficients of complex Ginzburg-Landau approximations to $n$-component reaction-diffusion systems, under various asymptotic limits, using the same approach as that adopted in the present paper. The formula for their cubic coefficient is equivalent to that for $C_3$ below, although derived in a different context. Also, because of our motivation to study super/sub-critical transitions, we here go up to order five. Moreover, in keeping with the aims of this journal, we provide all the details of the calculation and its implementation in computer algebra software.}
    
    The rest of the paper is outlined as follows. In Sec.~\ref{sec:main}, we provide a general form for the coefficients of the normal form of Turing bifurcation for an $n$-component reaction-diffusion system, expressed in the form of Theorem \evs{\ref{th:main}}, which is proved in Sec.~\ref{sec:proof}. \evs{We also show how to construct the additional coefficients of the extended fifth-order normal form.} Note that the proof is constructive, and hence is amenable to automatic evaluation for particular examples. Hence, Sec.~\ref{sec:comp} provides brief details of a numerical algorithm for the computation of these coefficients for an arbitrary example system. Further details are given in an appendix and the code can be found in a GitHub repository. Finally, Sec.~\ref{sec:examples} provides examples, illustrating cases with; cross-diffusion, higher-order spatial derivatives, and a 4-component system.

\section{Main result} \label{sec:main}
    Consider a general $n$-component reaction-diffusion system written in the form
    \begin{align}
    	\partial_t \mbf u &= \mbf f(\mbf u, \varepsilon) + \mathbb D(\varepsilon) \, \partial_{xx} \mbf u, \label{geneq}
    \end{align}
    \evs{for a vector function $\mbf u(x, t) \in \mathbb R^n$  with $x \in \mathcal D \subset \mathbb R$,} where $\mbf f\in \mathcal C^d$, for $d\geq 5$, $\varepsilon\in \mathbb R$ is a parameter of the system, and $\mathbb D(\varepsilon)$ is a positive definite diffusion matrix. \evs{We do not explicitly state boundary conditions to \eqref{geneq} as the result is valid when you consider either periodic or homogeneous Neumann boundary conditions.} We assume that $\mbf P = \mbf P(\varepsilon)$ is an isolated homogeneous steady state of \eqref{geneq}. Let us further denote the Jacobian matrix of the spatially-homogeneous system at $\mbf P$ by
	\begin{align*}
		\jac \mbf f(\mbf P, \varepsilon) = \begin{pmatrix}
			\frac{\partial f_i}{\partial u_j}(\mbf P, \varepsilon)
		\end{pmatrix}_{1\leq i, j\leq n},
	\end{align*}
    the diffusion matrix by
	\begin{align*}
		\mathbb D(\varepsilon) = \left(D_{i, j}(\varepsilon)\right)_{1\leq i, j\leq n},
	\end{align*}
    and define the following multilinear, symmetric, vector functions:
	\begin{align*}
		\mbf F_{1,1}(\mbf a) &= \sum_{1\leq p\leq n} a_p \, \left. \frac{\partial}{\partial \varepsilon} \left(\left. \frac{\partial \mbf f}{\partial u_p} \right|_{(\mbf P(\varepsilon),\varepsilon)}\right) \right|_{\varepsilon^*},
		\s 
		\mbf F_2(\mbf a,\mbf b) &= \frac{1}{2!} \left(\sum_{1\leq p,q\leq n} a_p b_q \, \left. \frac{\partial^2 \mbf f}{\partial u_p \partial u_q}\right|_{(\mbf P,\varepsilon^*)}\right),
		\s 
		\mbf F_3(\mbf a,\mbf b,\mbf c) &= \frac{1}{3!}\left(\sum_{1\leq p,q,r\leq n} a_p b_q c_r \, \left. \frac{\partial^3 \mbf f}{\partial u_p \partial u_q \partial u_r}\right|_{(\mbf P,\varepsilon^*)}\right),
		\s 
		\mbf F_4(\mbf a,\mbf b,\mbf c,\mbf d) &= \frac{1}{4!} \left(\sum_{1\leq p,q,r,s\leq n} a_p b_q c_r d_s \, \left. \frac{\partial^4 \mbf f}{\partial u_p \partial u_q \partial u_r \partial u_s}\right|_{(\mbf P,\varepsilon^*)}\right),
		\s 
		\mbf F_5(\mbf a,\mbf b,\mbf c,\mbf d,\mbf e) &= \frac{1}{5!} \left(\sum_{1\leq p,q,r,s,t\leq n} a_p b_q c_r d_s e_t \, \left .\frac{\partial^5 \mbf f}{\partial u_p \partial u_q \partial u_r \partial u_s \partial u_t} \right|_{(\mbf P,\varepsilon^*)}\right),
	\end{align*}
	for $\mbf a = \left(a_1, \ldots, a_n\right)^\intercal, \mbf b = \left(b_1, \ldots, b_n\right)^\intercal, \mbf c = \left(c_1, \ldots, c_n\right)^\intercal, \mbf d = \left(d_1, \ldots, d_n\right)^\intercal, \mbf e = \left(e_1, \ldots, e_n\right)^\intercal\in \mathbb R^n$, and $\varepsilon^*\in \mathbb R$.

    The purpose of this article is to provide a systematic method for determining the criticality \arc{of} a Turing bifurcation, and more generally for the calculation of normal form coefficients. We consider a simple form of the normal form for the amplitude equation that can be derived under the usual assumptions of weakly nonlocal approximation in the neighbourhood of the bifurcation\arc{. Assuming that the bifurcation occurs at the parameter value $\varepsilon = 0$, then the normal form that determines the criticality of the small-amplitude periodic patterns can be written in the form (see e.g.~\cite[Sec.6.1.3.1]{Meron})}
    \begin{align}
        \partial_t A &= C_{1, 1} \, A \, \varepsilon + C_3 \, |A|^2 A + C_5 \, |A|^4 A \: + \mathcal O\left(|A|^6, \varepsilon^2 |A|, \varepsilon \, |A|^2\right). \label{eq:NFeqn}
    \end{align}
    \arc{Note that if we wish to consider the weakly nonlinear dynamics of patterns in a neighbourhood of the bifurcation point, then there are additional nonlinear terms involving derivatives with respect to a long spatial variable $X$ and slow time variable $T$. The resulting Ginzburg-Landau type equation is given in \eqref{eq:CGL} below, and the derivation of the additional nonlinear terms is dealt with in Sec.~\ref{sec:3.6}.}
    
    Our main result can be stated as follows, in which we use the notation `$\ker$' to represent the kernel of a linear operator, and `$\range$' to represent its image (or range).

	\begin{theorem}\label{th:main}
        Let $\mbf P = \mbf P(\varepsilon)$ be a homogeneous steady state of \eqref{geneq}. \evs{Suppose} there exists a \textit{critical wavenumber} $\hat k > 0$, and $\varepsilon^*\in \mathbb R$ such that $\ker \left(\jac \mbf f \left(\mbf P, \varepsilon^*\right) - \hat k^2 \, \mathbb D\left(\varepsilon^*\right)\right) \cap \, \arc{\range} \left(\jac \mbf f \left(\mbf P, \varepsilon^*\right) - \hat k^2 \, \mathbb D\left(\varepsilon^*\right)\right) = \{\mbf 0\}$, $\dim \left(\ker \left(\jac \mbf f \left(\mbf P,\varepsilon^*\right)- \hat k^2 \, \mathbb D\left(\varepsilon^*\right)\right)\right)=1$, while $\ker \left(\jac \mbf f \left(\mbf P,\varepsilon^* \right)- k^2 \, \mathbb D\left(\varepsilon^*\right)\right) = \{\mbf 0\}$ for all $k\neq \hat k$. 
		\begin{align*}
			C_{1,1}&:=\frac{1}{\bs \psi_1^{[0]}\cdot \bs \phi_1^{[1]}} \, \bs \psi_1^{[0]} \cdot \left( \mbf F_{1,1}\left(\bs \phi_1^{[1]}\right) - \hat k^2 \, \frac{\dd \mathbb D}{\dd \varepsilon}\left(\varepsilon^*\right) \, \bs \phi_1^{[1]} \right)\neq 0,
		\end{align*}
		and
		\begin{align*}
			C_3\coloneqq \frac{1}{\bs \psi_1^{[0]} \cdot \bs \phi_1^{[1]}} \, \bs \psi_1^{[0]} \cdot \left(2 \, \mbf F_2\left(\bs \phi_1^{[1]}, 2 \, \evs{\mbf W}_0^{[2]} + \evs{\mbf W}_2^{[2]}\right) + 3 \, \mbf F_3\left(\bs \phi_1^{[1]}, \bs \phi_1^{[1]}, \bs \phi_1^{[1]}\right) \right)\neq 0,
		\end{align*}
		where 
		\begin{align*}
			\bs \phi_1^{[1]} &\in \ker(\jac \mbf f \left(\mbf P,\varepsilon^*\right)-\hat k^2 \, \mathbb D\left(\varepsilon^*\right))\setminus \{\mbf 0\},
			\s 
			\bs \psi_1^{[0]} &\in \ker(\jac \mbf f \left(\mbf P,\varepsilon^*\right)^\intercal-\hat k^2 \, \mathbb D\left(\varepsilon^*\right)^\intercal)\setminus \{\mbf 0\},
			\s 
			\jac \mbf f\left(\mbf P, \varepsilon^*\right) \, \evs{\mbf W}_0^{[2]} &= - \mbf F_2\left(\bs \phi_1^{[1]}, \bs \phi_1^{[1]}\right),
			\s 
			\left(\jac \mbf f\left(\mbf P,\varepsilon^*\right) - 4 \, \hat k^2 \, \mathbb D\left(\varepsilon^*\right)\right) \evs{\mbf W}_2^{[2]} &= -\mbf F_2\left(\bs \phi_1^{[1]}, \bs \phi_1^{[1]}\right),
		\end{align*}
		then $\mbf P$ goes through a Turing bifurcation which is subcritical if $C_3>0$ and supercritical if $C_3<0$.
		
        Furthermore, if $C_3=0$, then the bifurcation is of codimension-two, and the coefficient $C_5$ is given by
		\begin{align*}
			C_5 := \frac{1}{\bs \psi_1^{[0]} \cdot \bs \phi_1^{[1]}} \, \bs \psi_1^{[0]}\cdot \left(-3 \, C_3 \, \mbf W_1^{[3]} + 2 \, \mbf F_2\left(\bs \phi_1^{[1]}, 2 \, \mbf W_0^{[4]} + \mbf W_2^{[4]}\right) + 4 \, \mbf F_2\left(\mbf W_0^{[2]}, \mbf W_1^{[3]}\right)\right.
    		\s 
    		+ 2 \, \mbf F_2\left(\mbf W_2^{[2]},\mbf W_1^{[3]} + \mbf W_3^{[3]}\right) + 3 \, \mbf F_3\left(\bs \phi_1^{[1]}, \bs \phi_1^{[1]}, 3 \, \mbf W_1^{[3]} + \mbf W_3^{[3]}\right)
    		\s 
    		+ 12 \, \mbf F_3\left(\bs \phi_1^{[1]}, \mbf W_0^{[2]}, \mbf W_0^{[2]} + \mbf W_2^{[2]}\right) + 6 \, \mbf F_3\left(\bs \phi_1^{[1]}, \mbf W_2^{[2]}, \mbf W_2^{[2]}\right)
    		\s 
    		\left. + 8 \, \mbf F_4\left(\bs \phi_1^{[1]}, \bs \phi_1^{[1]}, \bs \phi_1^{[1]}, 3 \, \mbf W_0^{[2]} + 2 \, \mbf W_2^{[2]}\right) + 10 \, \mbf F_5\left(\bs \phi_1^{[1]}, \bs \phi_1^{[1]}, \bs \phi_1^{[1]}, \bs \phi_1^{[1]}, \bs \phi_1^{[1]}\right)\right),
		\end{align*}
		\evs{where}
		\begin{align*}
			\left(\jac \mbf f\left(\mbf P, \varepsilon^*\right) - \hat k^2 \, \mathbb D\left(\varepsilon^*\right)\right)\evs{\mbf W}_1^{[3]} = C_3 \, \bs \phi_1^{[1]} - 2 \, \mbf F_2\left(\bs \phi_1^{[1]}, 2 \, \evs{\mbf W}_0^{[2]} + \evs{\mbf W}_2^{[2]}\right) - 3 \, \mbf F_3\left(\bs \phi_1^{[1]},\bs \phi_1^{[1]},\bs \phi_1^{[1]}\right),
		\end{align*}
		\begin{align*}
			\left(\jac \mbf f\left(\mbf P,\varepsilon^*\right)-9 \, \hat k^2 \, \mathbb D\left(\varepsilon^*\right)\right)\evs{\mbf W}_3^{[3]} = - 2 \, \mbf F_2\left(\bs \phi_1^{[1]}, \evs{\mbf W}_2^{[2]}\right) - \mbf F_3\left(\bs \phi_1^{[1]}, \bs \phi_1^{[1]}, \bs \phi_1^{[1]}\right),
		\end{align*}
		\begin{multline*}
			\jac \mbf f\left(\mbf P,\varepsilon^*\right) \, \evs{\mbf W}_0^{[4]} = 2 \, C_3 \, \evs{\mbf W}_0^{[2]} - 2 \, \mbf F_2\left(\evs{\mbf W}_0^{[2]}, \evs{\mbf W}_0^{[2]}\right) - \mbf F_2\left(\evs{\mbf W}_2^{[2]}, \evs{\mbf W}_2^{[2]}\right) - 2 \, \mbf F_2\left(\bs \phi_1^{[1]}, \evs{\mbf W}_1^{[3]}\right)
            \s
            - 3 \, \mbf F_3\left(\bs \phi_1^{[1]}, \bs \phi_1^{[1]}, 2 \, \evs{\mbf W}_0^{[2]} + \evs{\mbf W}_2^{[2]}\right) - 3 \, \mbf F_4\left(\bs \phi_1^{[1]}, \bs \phi_1^{[1]}, \bs \phi_1^{[1]}, \bs \phi_1^{[1]}\right),
		\end{multline*}
        \evs{and}
		\begin{multline*}
			\left(\jac \mbf f\left(\mbf P,\varepsilon^*\right)-4 \, \hat k^2 \, \mathbb D\left(\varepsilon^*\right)\right)\evs{\mbf W}_2^{[4]} = 2 \, C_3 \, \evs{\mbf W}_2^{[2]} - 4 \, \mbf F_2\left(\evs{\mbf W}_0^{[2]}, \evs{\mbf W}_2^{[2]}\right) - 2 \, \mbf F_2\left(\bs \phi_1^{[1]}, \evs{\mbf W}_1^{[3]} + \evs{\mbf W}_3^{[3]}\right)
    		\s 
    		- 6 \, \mbf F_3\left(\bs \phi_1^{[1]}, \bs \phi_1^{[1]}, \evs{\mbf W}_0^{[2]} + \evs{\mbf W}_2^{[2]}\right) - 4 \, \mbf F_4\left(\bs \phi_1^{[1]}, \bs \phi_1^{[1]}, \bs \phi_1^{[1]}, \bs \phi_1^{[1]}\right).
		\end{multline*}
	\end{theorem}

\section{Proof of the main result}\label{sec:proof}        
    To prove the theorem, without loss of generality, we consider that $\mbf P = \mbf 0$ for all $\varepsilon\geq 0$. If this was not the case, we could always perform the following change of variables:
	\begin{equation*}
		\left(u_1, \ldots, u_n\right)^\intercal\mapsto \left(u_1, \ldots, u_n\right)^\intercal - \mbf P(\varepsilon),
	\end{equation*}
    which translates $\mbf P(\varepsilon)$ to the origin. Moreover, without loss of generality, we will assume that $\varepsilon^* = 0$. Again, if this was not the case and the conditions that $\jac \mbf f \left(\mbf 0, \varepsilon^*\right) - \hat k^2 \, \mathbb D\left(\varepsilon^*\right)$ must fulfill are satisfied when $\varepsilon = \varepsilon^*\neq 0$, then we could define a new parameter $\hat{\varepsilon} = \varepsilon - \varepsilon^*$ and our analysis will be valid for that system with the new parameter $\hat \varepsilon$. 
    	
    To simplify notation, we shall drop explicit $\varepsilon$ dependence by assuming, unless otherwise stated, that all functions are evaluated at $\varepsilon = 0$. We shall consider changes in this parameter in the last part of the proof. Also, from here on, we shall write $k$ instead of $\hat k$ for the \textit{critical wavenumber} stated in the theorem, under the assumption that all relevant quantities are evaluated at $k = \hat k$ unless otherwise stated. 
    	
    Under the hypotheses of Theorem \ref{th:main}, we can expand the vector function $\mbf f = \left(f_1, \ldots, f_n\right)^\intercal$ about $\mbf 0$ using a Taylor series with respect to $\mbf u = \left(u_1, \ldots, u_n\right)^\intercal$ as follows:
	\begin{align*}
		\mbf f(\mbf u) = \sum_{p = 1}^d \frac{1}{p!}\left(\sum_{m=1}^n u_m \frac{\partial }{\partial u_m} \right)^p \mbf f(\mbf 0) +\mathcal O\left(\evs{\norm{\mbf u}^{d + 1}}\right),
	\end{align*}
    where $\mbf 0$ is the $n$-component zero vector. Now, if we assume that the origin fulfills the conditions stated in the theorem, then we can build a normal form for the amplitude equation by means of the following ansatz:
	\begin{align*}
		\mbf u = A \, e^{ikx} \, \bs \phi_1^{[1]} + c.c.,
	\end{align*}
    where \evs{$c.c.$ stands for the complex conjugate of the expression to the left of it}, $\norm{\bs \phi_1^{[1]}} = 1$, $A = A(t)\in \mathbb C$ is a complex variable, and $|A|$ is the amplitude of a spatially periodic solution arising from the origin at the bifurcation point.
    
    In particular, at the bifurcation point when $\varepsilon = 0$, we want $A$ to solve an equation of the following form:
	\begin{align}
		\partial_t A = h^{[1]}(A) + h^{[2]}(A) + \ldots, \label{geneqA}
	\end{align}
    where each function $h^{[i]}(A)$, for $i = 1, 2, \ldots$ corresponds to a linear function in $A^i$. For that purpose, we need to find a suitable change of coordinates
	\begin{align*}
		\mbf u = \evs{\mbf u}^{[1]}(A) + \evs{\mbf u}^{[2]}(A) + \ldots,
	\end{align*}
    such that $A$ solves an equation of the form \eqref{geneqA}. In particular, we are interested in doing this up to order five. We shall proceed order by order. 
    	
\subsection{Order 1}
    Up to first order\evs{,} \eqref{geneq} becomes
	\begin{align*}
		\partial_A \evs{\mbf u}^{[1]} \,  h^{[1]} + \mbox{conj.} &= \left(\jac \mbf f(\mbf 0) + \mathbb D \, \partial_{xx}\right)\evs{\mbf u}^{[1]},
	\end{align*}
    where $\mbox{conj.}$ stands for the conjugate version of the term to the left, which is not necessarily its complex conjugate. For instance, $\partial_A \evs{\mbf u^{[1]}} \, h_1^{\evs{[1]}} + \mbox{conj.} = \partial_A \evs{\mbf u^{[1]}} \, h_1^{\evs{[1]}} + \partial_{\bar A} \evs{\mbf u^{[1]}} \, \overline{h_1^{\evs{[1]}}}$.
    
	Therefore, if we choose $h^{[1]} = 0$, then we will need to solve 
	\begin{align}
		\left(\jac \mbf f(\mbf 0) + \mathbb D \, \partial_{xx}\right)\evs{\mbf u}^{[1]} &= 0. \label{firstordereq}
	\end{align}
    As we know that $\dim\left(\ker \left(\jac \mbf f(\mbf 0) - k^2 \, \mathbb D \right)\right) = 1$, then there exists $\bs \phi_1^{[1]} \in \ker \left(\jac \mbf f(\mbf 0) - k^2 \, \mathbb D\right) \setminus \{\mbf 0\}$, such that
	\begin{align*}
		\ker \left(\jac \mbf f(\mbf 0) - k^2 \, \mathbb D \right) = \mbox {span}\left(\bs \phi_1^{[1]}\right).
	\end{align*}
    With this, we can choose a solution to \eqref{firstordereq} of the form
	\begin{align*}
		\evs{\mbf u}^{[1]} = A \, e^{ikx}\bs \phi_1^{[1]} + c.c.
	\end{align*}

\subsection{Order 2}
    At second order, \eqref{geneq} is given by 
	\begin{align}
		\partial_A \evs{\mbf u}^{[1]} \, h^{[2]} + \mbox{conj.} = \left(\jac \mbf f(\mbf 0) + \mathbb D \, \partial_{xx}\right)\evs{\mbf u}^{[2]} + \mbf F_2\left(\evs{\mbf u}^{[1]}, \evs{\mbf u}^{[1]}\right) \label{secondorder}
	\end{align}
    Then, as $\partial_A \evs{\mbf u}^{[1]} \in \ker\left(\jac \mbf f(\mbf 0) - k^2 \, \mathbb D\right)$, $\ker\left(\jac \mbf f(\mbf 0) - k^2 \, \mathbb D\right) \cap \, \evs{\range}\left(\jac \mbf f(\mbf 0) - k^2 \, \mathbb D\right) = \left\{\mbf 0\right\}$ and there are no secular terms in the right-hand side of \eqref{secondorder}, there is no way that we can solve this equation if $h^{[2]}\neq 0$. Therefore, we must set $h^{[2]} = 0$, which gives
	\begin{align}
		\left(\jac \mbf f(\mbf 0) + \mathbb D \, \partial_{xx}\right)\evs{\mbf u}^{[2]} = - \left(|A|^2 + A^2 \, e^{2ikx}\right) \, \mbf F_2\left(\bs \phi_1^{[1]}, \bs \phi_1^{[1]}\right) + c.c. \label{secondordergeneral}
	\end{align}
    Note that \eqref{secondordergeneral} has infinitely many solutions but only one that is orthogonal to $\ker\left(\jac \mbf f(\mbf 0)+\mathbb D \, \partial_{xx}\right)$, namely 
	\begin{align*}
		\evs{\mbf u}^{[2]}=|A|^2 \, \evs{\mbf W}_0^{[2]} + A^2 \, e^{2ikx} \, \evs{\mbf W}_2^{[2]} + c.c.
	\end{align*}
    In particular, note that \eqref{secondordergeneral} is a linear equation. Thus, in order to find $\evs{\mbf W}_0^{[2]}$ and $\evs{\mbf W}_2^{[2]}$, we can use the principle of linear superposition. In particular, we have
	\begin{align*}
		\jac \mbf f(\mbf 0) \, \evs{\mbf W}_0^{[2]} = - \mbf F_2\left(\bs \phi_1^{[1]}, \bs \phi_1^{[1]}\right),
	\end{align*}
	and
	\begin{align*}
		\left(\jac \mbf f(\mbf 0) - 4 \, k^2 \, \mathbb D\right) \evs{\mbf W}_2^{[2]} = - \mbf F_2\left(\bs \phi_1^{[1]}, \bs \phi_1^{[1]}\right).
	\end{align*}
    \evs{N}ote that both of these linear systems have a unique solution, \evs{as} $\ker\left(\jac \mbf f(\mbf 0) - \ell^2 \, \mathbb D\right) = \{\mbf 0\}$ for all $\ell\neq k$ and $\ell^2 > 0$.
    	
\subsection{Order 3}
    At third order, \eqref{geneq} is given by
	\begin{align}
		\partial_A \evs{\mbf u}^{[1]} \, h^{[3]} + \mbox{conj.} = \left(\jac \mbf f(\mbf 0) + \mathbb D \, \partial_{xx}\right)\evs{\mbf u}^{[3]} + 2 \, \mbf F_2\left(\evs{\mbf u}^{[1]}, \evs{\mbf u}^{[2]}\right) + \mbf F_3\left(\evs{\mbf u}^{[1]}, \evs{\mbf u}^{[1]}, \evs{\mbf u}^{[1]}\right). \label{thirdordereq}
	\end{align}
    Now, using the definitions of $\evs{\mbf u}^{[1]}$ and $\evs{\mbf u}^{[2]}$, we can see that \eqref{thirdordereq} can be expanded as
	\begin{multline}
		\left(\jac \mbf f(\mbf 0) + \mathbb D \, \partial_{xx}\right)\evs{\mbf u}^{[3]} = e^{ikx} \, \bs \phi_1^{[1]} \, h^{[3]}
        \\
        - 2 \, \abs{A}^2 A \, e^{ikx} \, \mbf F_2\left(\bs \phi_1^{[1]}, 2 \, \evs{\mbf W}_0^{[2]} + \evs{\mbf W}_2^{[2]}\right) \evs{- 3 \, \abs{A}^2 A \, e^{ikx} \, \mbf F_3\left(\bs \phi_1^{[1]},\bs \phi_1^{[1]},\bs \phi_1^{[1]}\right)}
        \\
        \evs{- 2 \, A^3 \, e^{3ikx} \, \mbf F_2\left(\bs \phi_1^{[1]}, \evs{\mbf W}_2^{[2]}\right)} - A^3 \, e^{3ikx} \, \mbf F_3\left(\bs \phi_1^{[1]},\bs \phi_1^{[1]},\bs \phi_1^{[1]}\right) + c.c. \label{generalthirdorder}
	\end{multline}
    Note that the right-hand side of \eqref{generalthirdorder} will, in general, contain secular terms \evs{as} $\jac \mbf f(\mbf 0)-k^2 \, \mathbb D$ is non-invertible. Using the Fredholm alternative, we have that
	\begin{align*}
		\left(\evs{\range}\left(\jac \mbf f(\mbf 0)-k^2 \, \mathbb D\right) \right)^\perp =\ker\left(\left(\jac \mbf f(\mbf 0)-k^2 \, \mathbb D \right)^* \right).
	\end{align*}
	Therefore, to know the image of $\jac \mbf f(\mbf 0)-k^2 \, \mathbb D$, we need to find the kernel of its adjoint, which is spanned by vectors $\bs \psi_1^{[0]}\neq \mbf 0$ such that
	\begin{align*}
		\left(\jac \mbf f(\mbf 0)-k^2 \, \mathbb D\right)^*\bs \psi_1^{[0]} &=\mbf 0
		\s 
		\iff \left(\jac \mbf f(\mbf 0)^\intercal-k^2 \, \mathbb D^\intercal\right)\bs \psi_1^{[0]} &=\mbf 0.
	\end{align*}
    This implies that $\ker\left(\left(\jac \mbf f(\mbf 0)+\mathbb D \, \partial_{xx}\right)^*\right)$ is spanned by the vector functions $\bs{\hat \psi}_\pm$ given by
	\begin{align*}
		\bs{\hat \psi}_\pm=e^{\pm ikx} \, \bs \psi_1^{[0]},
	\end{align*}
    Thus, a sufficient condition for the solvability of the third-order equation is that the inner product between the right-hand side of \eqref{generalthirdorder} and each of the functions $\bs{\hat \psi}_\pm$ must be equal to zero, which ensures that the right-hand side of it belongs to $\evs{\range}\left(\jac \mbf f(\mbf 0)+\mathbb D \, \partial_{xx}\right)$.
    
    A natural inner product, that we shall adopt here is given by 
	\begin{align*}
		\innerp{\mbf a \, f(x)}{\mbf b \, g(x)} &= \frac{k}{2 \, \pi}\int_0^{\frac{2\pi}{k}} f(x) \, \overline{g(x)} \, \dd x \, \mbf a \cdot \mbf b,
	\end{align*}
    where $a,b\in \mathbb R^n$, and $\cdot$ is used to denote the usual inner-product between vectors in that space.

    We can now compute
	\begin{align*}
		\innerp{\left(\jac \mbf f(\mbf 0) + \mathbb D \, \partial_{xx}\right)\mbf W^{[3]}}{\bs{\hat \psi}_+} = \innerp{e^{ikx} \, h^{[3]} \, \bs \phi_1^{[1]}}{e^{ikx} \, \bs \psi_1^{[0]}}
		\s 
		- 2 \, |A|^2 A \innerp{e^{ikx} \, \mbf F_2\left(\bs \phi_1^{[1]}, 2 \, \evs{\mbf W}_0^{[2]} + \evs{\mbf W}_2^{[2]}\right)}{e^{ikx} \, \bs \psi_1^{[0]}}
		\s 
		- 3 \, \abs{A}^2 A \innerp{e^{ikx} \, \mbf F_3\left(\bs \phi_1^{[1]}, \bs \phi_1^{[1]}, \bs \phi_1^{[1]}\right)}{e^{ikx} \, \bs \psi_1^{[0]}}.
	\end{align*}
    Upon setting this expression to zero, we find that $h^{[3]}(A) = C_3 \, |A|^2 A$, where
	\begin{align}
		C_3 = \frac{1}{\bs \psi_1^{[0]} \cdot \bs \phi_1^{[1]}} \, \bs \psi_1^{[0]} \cdot \left(2 \, \mbf F_2\left(\bs \phi_1^{[1]}, 2 \, \evs{\mbf W}_0^{[2]} + \evs{\mbf W}_2^{[2]}\right) + 3 \, \mbf F_3\left(\bs \phi_1^{[1]}, \bs \phi_1^{[1]}, \bs \phi_1^{[1]}\right) \right). \label{eq:C3}
	\end{align}
	\textbf{Remark.} Note that the term in the denominator of $C_3$ can be shown to be nonzero. This is because for a linear operator $L$ with a 1-dimensional kernel, \evs{if} $\mbf 0 \neq \bs \psi\in \ker\left(L^*\right)=\left(\evs{\range}(L)\right)^\perp$, then
	\begin{align*}
		\left \langle \bs \phi, \bs \psi \right \rangle =0 \qquad \iff \qquad \bs \phi \in \evs{\range}(L).
	\end{align*}
    In our case, we have that $\bs \phi_1^{[1]}\in \ker \left(\jac \mbf f(\mbf 0)-k^2 \, \mathbb D\right)\setminus \{\mbf 0\}$ and $\bs \psi_1^{[0]}\in \ker \left(\left(\jac \mbf f(\mbf 0)-k^2 \, \mathbb D\right)^*\right)\setminus \{\mbf 0\}$ but, since $\ker \left(\jac \mbf f(\mbf 0)-k^2 \, \mathbb D\right) \cap \, \evs{\range} \left(\jac \mbf f(\mbf 0)-k^2 \, \mathbb D\right)=\{\mbf 0\}$, then $\bs \psi_1^{[0]} \cdot \bs \phi_1^{[1]}\neq 0$.

    Now, in order to continue coordinate transformations up to order 5, we need to obtain $\evs{\mbf u}^{[3]}$ explicitly. In particular, if we substitute $h^{[3]}$ into \eqref{generalthirdorder}, then we know that the equation will have a solution that can be written in the form 
	\begin{align*}
		\evs{\mbf u}^{[3]} = |A|^2 A \, e^{ikx} \, \evs{\mbf W}_1^{[3]} + A^3 \, e^{3ikx} \, \evs{\mbf W}_3^{[3]} + c.c.
	\end{align*}
    Then, because \eqref{generalthirdorder} is a linear system, we can proceed to determine $\evs{\mbf W}_1^{[3]}$ and $\evs{\mbf W}_3^{[3]}$ by substitution of $\evs{\mbf u}^{[3]}$ term by term. Upon doing so, we obtain the following equations:
	\begin{align*}
		\left(\jac \mbf f(\mbf 0) - k^2 \, \mathbb D\right)\evs{\mbf W}_1^{[3]} = C_3 \, \bs \phi_1^{[1]} - 2 \, \mbf F_2\left(\bs \phi_1^{[1]}, 2 \, \evs{\mbf W}_0^{[2]} + \evs{\mbf W}_2^{[2]}\right) - 3 \, \mbf F_3\left(\bs \phi_1^{[1]}, \bs \phi_1^{[1]}, \bs \phi_1^{[1]}\right),
	\end{align*}
    and
	\begin{align*}
		\left(\jac \mbf f(\mbf 0) - 9 \, k^2 \, \mathbb D\right)\evs{\mbf W}_3^{[3]} = - 2 \, \mbf F_2\left(\bs \phi_1^{[1]}, \evs{\mbf W}_2^{[2]}\right) - \mbf F_3\left(\bs \phi_1^{[1]}, \bs \phi_1^{[1]}, \bs \phi_1^{[1]}\right),
	\end{align*}
    respectively. Note that the equation for $\evs{\mbf W}_1^{[3]}$ has infinitely many solutions since the right-hand side of that equation belongs to the range of $\jac \mbf f(\mbf 0) - k^2 \, \mathbb D$ due to the solvability condition we used to determine the value of $C_3$, but that matrix is non\evs{-}invertible. To find a unique solution, a natural choice is to \evs{require} that 
    \begin{align*}
            \evs{\mbf W}_1^{[3]}\cdot \bs \phi_1^{[1]} = 0.
    \end{align*}
    This is consistent with the asymptotic expansion, as it assumes that all information in the direction of    
    $\bs \phi\evs{_1^{[1]}}$ was considered at first order.
    		
\subsection{Order 4}
    At fourth order, \eqref{geneq} is given by
	\begin{multline}
		\partial_A \evs{\mbf u}^{[1]} \, h^{[4]} + \partial_A \evs{\mbf u}^{[2]} \, h^{[3]} + \mbox{conj.} = \left(\jac \mbf f(\mbf 0) + \mathbb D \, \partial_{xx}\right)\evs{\mbf u}^{[4]} \evs{+ 2 \, \mbf F_2\left(\evs{\mbf u}^{[1]}, \evs{\mbf u}^{[3]}\right) + \mbf F_2\left(\evs{\mbf u}^{[2]}, \evs{\mbf u}^{[2]}\right)}
		\s 
		+ 3 \, \mbf F_3\left(\evs{\mbf u}^{[1]}, \evs{\mbf u}^{[1]}, \evs{\mbf u}^{[2]}\right) + \mbf F_4\left(\evs{\mbf u}^{[1]}, \evs{\mbf u}^{[1]}, \evs{\mbf u}^{[1]}, \evs{\mbf u}^{[1]}\right). \label{generalfourthorder}
	\end{multline}
    By the same argument used when developing the expression for order 2, we again need to enforce $h^{[4]}=0$. Now, note that
	\begin{align*}
        2 \, \mbf F_2\left(\evs{\mbf u}^{[1]}, \evs{\mbf u}^{[3]}\right) &= 2 \, \abs{A}^4 \, \mbf F_2\left(\bs \phi_1^{[1]}, \evs{\mbf W}_1^{[3]}\right) + 2 \, |A|^2 A^2 \, e^{2ikx} \, \mbf F_2\left(\bs \phi_1^{[1]}, \evs{\mbf W}_1^{[3]} + \evs{\mbf W}_3^{[3]}\right)
		\s 
		&\quad + 2 \, A^4 \, e^{4ikx} \, \mbf F_2\left(\bs \phi_1^{[1]}, \evs{\mbf W}_3^{[3]} \right) + c.c.,
        \s
		\mbf F_2\left(\evs{\mbf u}^{[2]}, \evs{\mbf u}^{[2]}\right) &= |A|^4 \left(2 \, \mbf F_2\left(\evs{\mbf W}_0^{[2]}, \evs{\mbf W}_0^{[2]}\right) + \mbf F_2\left(\evs{\mbf W}_2^{[2]}, \evs{\mbf W}_2^{[2]}\right)\right) + 4 \, |A|^2 A^2 \, e^{2ikx} \, \mbf F_2\left(\evs{\mbf W}_0^{[2]}, \evs{\mbf W}_2^{[2]}\right)
		\s 
		& \quad + A^4 \, e^{4ikx} \, \mbf F_2\left(\evs{\mbf W}_2^{[2]}, \evs{\mbf W}_2^{[2]}\right) + c.c.,
	\end{align*}
	\begin{align*}
		3 \, \mbf F_3\left(\evs{\mbf u}^{[1]}, \evs{\mbf u}^{[1]}, \evs{\mbf u}^{[2]}\right) &= 3 \, \abs{A}^4 \, \mbf F_3\left(\bs \phi_1^{[1]}, \bs \phi_1^{[1]}, 2 \, \evs{\mbf W}_0^{[2]} + \evs{\mbf W}_2^{[2]}\right)
        \\
        & \quad + 6 \, |A|^2 A^2 \, e^{2ikx} \, \mbf F_3\left(\bs \phi_1^{[1]}, \bs \phi_1^{[1]}, \evs{\mbf W}_0^{[2]} + \evs{\mbf W}_2^{[2]}\right)
        \s  
        &\quad + 3 \, A^4 \, e^{4ikx} \, \mbf F_3\left(\bs \phi_1^{[1]}, \bs \phi_1^{[1]}, \evs{\mbf W}_2^{[2]}\right) \evs{+ c.c.},
	  \s
		\mbf F_4\left(\evs{\mbf u}^{[1]}, \evs{\mbf u}^{[1]}, \evs{\mbf u}^{[1]}, \evs{\mbf u}^{[1]}\right) &= 3 \, \abs{A}^4 \, \mbf F_4\left(\bs \phi_1^{[1]}, \bs \phi_1^{[1]}, \bs \phi_1^{[1]}, \bs \phi_1^{[1]}\right)
        \\
        &\quad + 4 \, |A|^2 A^2 \, e^{2ikx} \, \mbf F_4\left(\bs \phi_1^{[1]}, \bs \phi_1^{[1]}, \bs \phi_1^{[1]}, \bs \phi_1^{[1]}\right)
		\s 
		& \quad + A^4 \, e^{4ikx} \, \mbf F_4\left(\bs \phi_1^{[1]}, \bs \phi_1^{[1]}, \bs \phi_1^{[1]}, \bs \phi_1^{[1]}\right) \evs{+ c.c.}.
	\end{align*}
    With this, we can write \eqref{generalfourthorder} as:
	\begin{multline*}
		\left(\jac \mbf f(\mbf 0) + \mathbb D \, \partial_{xx}\right)\evs{\mbf u}^{[4]} = 2 \, C_3 \, |A|^4 \, \evs{\mbf W}_0^{[2]} + 2 \, C_3 \, |A|^2 A^2 \, e^{2ikx} \, \evs{\mbf W}_2^{[2]}
		\s 
        \evs{- 2 \, \abs{A}^4 \, \mbf F_2\left(\bs \phi_1^{[1]}, \evs{\mbf W}_1^{[3]}\right) - 2 \, |A|^2 A^2 \, e^{2ikx} \, \mbf F_2\left(\bs \phi_1^{[1]}, \evs{\mbf W}_1^{[3]} + \evs{\mbf W}_3^{[3]}\right)}
        \s
        \evs{- |A|^4 \left(2 \, \mbf F_2\left(\evs{\mbf W}_0^{[2]}, \evs{\mbf W}_0^{[2]}\right) + \mbf F_2\left(\evs{\mbf W}_2^{[2]}, \evs{\mbf W}_2^{[2]}\right)\right) - 4 \, |A|^2 A^2 \, e^{2ikx} \, \mbf F_2\left(\evs{\mbf W}_0^{[2]}, \evs{\mbf W}_2^{[2]}\right)}
        \s 
        - 3 \, \abs{A}^4 \, \mbf F_3\left(\bs \phi_1^{[1]}, \bs \phi_1^{[1]}, 2 \, \evs{\mbf W}_0^{[2]} + \evs{\mbf W}_2^{[2]}\right) - 6 \, |A|^2 A^2 \, e^{2ikx} \, \mbf F_3\left(\bs \phi_1^{[1]}, \bs \phi_1^{[1]}, \evs{\mbf W}_0^{[2]} + \evs{\mbf W}_2^{[2]}\right)
        \s  
        - 3 \, \abs{A}^4 \, \mbf F_4\left(\bs \phi_1^{[1]}, \bs \phi_1^{[1]}, \bs \phi_1^{[1]}, \bs \phi_1^{[1]}\right) - 4 \, |A|^2 A^2 \, e^{2ikx} \, \mbf F_4\left(\bs \phi_1^{[1]}, \bs \phi_1^{[1]}, \bs \phi_1^{[1]}, \bs \phi_1^{[1]}\right) + \ldots + c.c.
	\end{multline*}
    \evs{where $\ldots$ represents terms that will not be used in this paper, as they do not produce any singularities up to order 5.}
    
    Once again, this equation has infinitely many solutions, but we are interested in the one that is orthogonal to $\ker\left(\jac \mbf f(\mbf 0) + \mathbb D \, \partial_{xx}\right)$. Thus, the solution we seek has the form
	\begin{align*}
		\evs{\mbf u}^{[4]} = |A|^4 \, \evs{\mbf W}_0^{[4]} + |A|^2 A^2 \, e^{2ikx} \, \evs{\mbf W}_2^{[4]} + \evs{\ldots} + c.c.,
	\end{align*}
    Finally, we can use the principle of linear superposition to find $\evs{\mbf W}_0^{[4]}$ and $\evs{\mbf W}_2^{[4]}$ by solving separately for each term of $\evs{\mbf u}^{[4]}$. \evs{Thus}, we obtain 
    \begin{multline*}
		\jac \mbf f(\mbf 0) \, \evs{\mbf W}_0^{[4]} = 2 \, C_3 \, \evs{\mbf W}_0^{[2]} \evs{- 2 \, \mbf F_2\left(\bs \phi_1^{[1]}, \evs{\mbf W}_1^{[3]}\right) - 2 \, \mbf F_2\left(\evs{\mbf W}_0^{[2]}, \evs{\mbf W}_0^{[2]}\right) - \mbf F_2\left(\evs{\mbf W}_2^{[2]}, \evs{\mbf W}_2^{[2]}\right)}
        \s
        - 3 \, \mbf F_3\left(\bs \phi_1^{[1]}, \bs \phi_1^{[1]}, 2 \, \evs{\mbf W}_0^{[2]} + \evs{\mbf W}_2^{[2]}\right) - 3 \, \mbf F_4\left(\bs \phi_1^{[1]}, \bs \phi_1^{[1]}, \bs \phi_1^{[1]}, \bs \phi_1^{[1]}\right),
    \end{multline*}
    \evs{and}
    \begin{multline*}
		\left(\jac \mbf f(\mbf 0) - 4 \, k^2 \, \mathbb D\right)\evs{\mbf W}_2^{[4]} = 2 \, C_3 \, \evs{\mbf W}_2^{[2]} \evs{- 2 \, \mbf F_2\left(\bs \phi_1^{[1]}, \evs{\mbf W}_1^{[3]} + \evs{\mbf W}_3^{[3]}\right) - 4 \, \mbf F_2\left(\evs{\mbf W}_0^{[2]}, \evs{\mbf W}_2^{[2]}\right)}
		\s 
		- 6 \, \mbf F_3\left(\bs \phi_1^{[1]}, \bs \phi_1^{[1]}, \evs{\mbf W}_0^{[2]} + \evs{\mbf W}_2^{[2]}\right) - 4 \, \mbf F_4\left(\bs \phi_1^{[1]}, \bs \phi_1^{[1]}, \bs \phi_1^{[1]}, \bs \phi_1^{[1]}\right)\evs{.}
    \end{multline*}
    With these \evs{vectors} defined in a unique way, we are ready to proceed to the final order we are interested in.

\subsection{Order 5}
    At fifth order, \eqref{geneq} is given by
	\begin{multline}
		\partial_A \evs{\mbf u}^{[1]} \, h^{[5]} + \partial_A \evs{\mbf u}^{[3]} \, h^{[3]} + \mbox{conj.} = \left(\jac \mbf f(\mbf 0)+\mathbb D \, \partial_{xx}\right) \evs{\mbf u}^{[5]} + 2 \, \mbf F_2\left(\evs{\mbf u}^{[1]}, \evs{\mbf u}^{[4]}\right) + 2 \, \mbf F_2\left(\evs{\mbf u}^{[2]}, \evs{\mbf u}^{[3]}\right)
		\s 
		+ 3 \, \mbf F_3\left(\evs{\mbf u}^{[1]}, \evs{\mbf u}^{[1]}, \evs{\mbf u}^{[3]}\right) + 3 \, \mbf F_3\left(\evs{\mbf u}^{[1]}, \evs{\mbf u}^{[2]}, \evs{\mbf u}^{[2]}\right)
		\s 
		+ 4 \, \mbf F_4\left(\evs{\mbf u}^{[1]}, \evs{\mbf u}^{[1]}, \evs{\mbf u}^{[1]}, \evs{\mbf u}^{[2]}\right) + \mbf F_5\left(\evs{\mbf u}^{[1]}, \evs{\mbf u}^{[1]}, \evs{\mbf u}^{[1]}, \evs{\mbf u}^{[1]}, \evs{\mbf u}^{[1]}\right). \label{generalfifthorder}
	\end{multline}
	Next, note that
	\begin{align*}
		2 \, \mbf F_2\left(\evs{\mbf u}^{[1]}, \evs{\mbf u}^{[4]}\right) &= 2 \, |A|^4 A \, e^{ikx} \, \mbf F_2\left(\bs \phi_1^{[1]}, 2 \, \evs{\mbf W}_0^{[4]} + \evs{\mbf W}_2^{[4]}\right) \evs{+ \ldots + c.c.},
		\s 
		2 \, \mbf F_2\left(\evs{\mbf u}^{[2]}, \evs{\mbf u}^{[3]}\right) &= 2 \, |A|^4 A \, e^{ikx} \left(2 \, \mbf F_2\left(\evs{\mbf W}_0^{[2]}, \evs{\mbf W}_1^{[3]}\right) + \mbf F_2\left(\evs{\mbf W}_2^{[2]}, \evs{\mbf W}_1^{[3]} + \evs{\mbf W}_3^{[3]}\right)\right) \evs{+ \ldots + c.c.},
	\end{align*}
	\begin{align*}
		3 \, \mbf F_3\left(\evs{\mbf u}^{[1]}, \evs{\mbf u}^{[1]}, \evs{\mbf u}^{[3]}\right) &= 3 \, \abs{A}^4 A \, e^{ikx} \, \mbf F_3\left(\bs \phi_1^{[1]}, \bs \phi_1^{[1]}, 3 \, \evs{\mbf W}_1^{[3]} + \evs{\mbf W}_3^{[3]}\right) \evs{+ \ldots + c.c.},
		\s 
		3 \, \mbf F_3\left(\evs{\mbf u}^{[1]}, \evs{\mbf u}^{[2]}, \evs{\mbf u}^{[2]}\right) &= 6 \, \abs{A}^4 A \, e^{ikx} \left(2 \, \mbf F_3\left(\bs \phi_1^{[1]}, \evs{\mbf W}_0^{[2]}, \evs{\mbf W}_0^{[2]} + \evs{\mbf W}_2^{[2]}\right) \right.
        \s
        & \quad \left. + \mbf F_3\left(\bs \phi_1^{[1]}, \evs{\mbf W}_2^{[2]}, \evs{\mbf W}_2^{[2]}\right)\right) \evs{+ \ldots + c.c.},
        \s
        4 \, \mbf F_4\left(\evs{\mbf u}^{[1]}, \evs{\mbf u}^{[1]}, \evs{\mbf u}^{[1]}, \evs{\mbf u}^{[2]}\right) &= 8 \, \abs{A}^4 A \, e^{ikx} \mbf F_4\left(\bs \phi_1^{[1]}, \bs \phi_1^{[1]}, \bs \phi_1^{[1]}, 3 \, \evs{\mbf W}_0^{[2]} + 2 \, \evs{\mbf W}_2^{[2]}\right) \evs{+ \ldots + c.c.},
        \s
        \mbf F_5\left(\evs{\mbf u}^{[1]}, \evs{\mbf u}^{[1]}, \evs{\mbf u}^{[1]}, \evs{\mbf u}^{[1]}, \evs{\mbf u}^{[1]}\right) &= 10 \, |A|^4 A \, e^{ikx} \, \mbf F_5\left(\bs \phi_1^{[1]}, \bs \phi_1^{[1]}, \bs \phi_1^{[1]}, \bs \phi_1^{[1]}, \bs \phi_1^{[1]}\right) \evs{+ \ldots + c.c.}
	\end{align*}
    Note that, in general, \eqref{generalfifthorder} will have secular terms and, to get rid of this problem, we will need to determine the coefficient of $h^{[5]}$, again, by using the inner product with $\bs{\hat \psi}_\pm$. Thus, we obtain
	\begin{multline*}
		\innerp{\left(\jac \mbf f(\mbf 0)+\mathbb D \, \partial_{xx}\right) \evs{\mbf u}^{[5]}}{\bs{\hat \psi}_\pm} = \innerp{e^{ikx} \, h^{[5]} \, \bs \phi_1^{[1]}}{e^{ikx} \, \bs \psi_1^{[0]}} + 3 \, C_3 \, |A|^4 A \innerp{e^{ikx} \, \evs{\mbf W}_1^{[3]}}{e^{ikx} \, \bs \psi_1^{[0]}}
		\s 
		- 2 \, |A|^4 A \innerp{e^{ikx} \, \mbf F_2\left(\bs \phi_1^{[1]}, 2 \, \evs{\mbf W}_0^{[4]} + \evs{\mbf W}_2^{[4]}\right)}{e^{ikx} \, \bs \psi_1^{[0]}}
		\s 
		- 4 \, |A|^4 A \innerp{e^{ikx} \, \mbf F_2\left(\evs{\mbf W}_0^{[2]}, \evs{\mbf W}_1^{[3]}\right)}{e^{ikx} \, \bs \psi_1^{[0]}}
		\s 
		- 2 \, |A|^4 A \innerp{e^{ikx} \, \mbf F_2\left(\evs{\mbf W}_2^{[2]}, \evs{\mbf W}_1^{[3]} + \evs{\mbf W}_3^{[3]}\right)}{e^{ikx} \, \bs \psi_1^{[0]}}
		\s 
		- 3 \, |A|^4 A \, \innerp{e^{ikx} \, \mbf F_3\left(\bs \phi_1^{[1]}, \bs \phi_1^{[1]}, 3 \, \evs{\mbf W}_1^{[3]} + \evs{\mbf W}_3^{[3]}\right)}{e^{ikx} \, \bs \psi_1^{[0]}}
		\s 
		- 12 \, |A|^4 A \, \innerp{e^{ikx} \, \mbf F_3\left(\bs \phi_1^{[1]}, \evs{\mbf W}_0^{[2]}, \evs{\mbf W}_0^{[2]} + \evs{\mbf W}_2^{[2]}\right)}{e^{ikx} \, \bs \psi_1^{[0]}}
		\s 
		- 6 \, |A|^4 A \innerp{e^{ikx} \mbf F_3\left(\bs \phi_1^{[1]}, \evs{\mbf W}_2^{[2]}, \evs{\mbf W}_2^{[2]}\right)}{e^{ikx} \, \bs \psi_1^{[0]}}
		\s 
		- 8 \, |A|^4 A \innerp{e^{ikx} \, \mbf F_4\left(\bs \phi_1^{[1]}, \bs \phi_1^{[1]}, \bs \phi_1^{[1]}, 3 \, \evs{\mbf W}_0^{[2]} + 2 \, \evs{\mbf W}_2^{[2]}\right)}{e^{ikx} \, \bs \psi_1^{[0]}}
		\s 
		- 10 \, |A|^4 A \, \innerp{e^{ikx} \, \mbf F_5\left(\bs \phi_1^{[1]}, \bs \phi_1^{[1]}, \bs \phi_1^{[1]}, \bs \phi_1^{[1]}, \bs \phi_1^{[1]}\right)}{e^{ikx} \, \bs \psi_1^{[0]}},
    \end{multline*}
    which implies that $h^{[5]}(A)=C_5 \, |A|^4 A$, where
	\begin{multline}
		C_5 = \frac{1}{\bs \psi_1^{[0]} \cdot \bs \phi_1^{[1]}} \, \bs \psi_1^{[0]}\cdot \left(-3 \, C_3 \, \evs{\mbf W}_1^{[3]} + 2 \, \mbf F_2\left(\bs \phi_1^{[1]}, 2 \, \evs{\mbf W}_0^{[4]} + \evs{\mbf W}_2^{[4]}\right) + 4 \, \mbf F_2\left(\evs{\mbf W}_0^{[2]}, \evs{\mbf W}_1^{[3]}\right)\right.
		\s 
		+ 2 \, \mbf F_2\left(\evs{\mbf W}_2^{[2]}, \evs{\mbf W}_1^{[3]} + \evs{\mbf W}_3^{[3]}\right) + 3 \, \mbf F_3\left(\bs \phi_1^{[1]}, \bs \phi_1^{[1]}, 3 \, \evs{\mbf W}_1^{[3]} + \evs{\mbf W}_3^{[3]}\right)
		\s 
		+ 12 \, \mbf F_3\left(\bs \phi_1^{[1]}, \evs{\mbf W}_0^{[2]}, \evs{\mbf W}_0^{[2]} + \evs{\mbf W}_2^{[2]}\right) + 6 \, \mbf F_3\left(\bs \phi_1^{[1]}, \evs{\mbf W}_2^{[2]}, \evs{\mbf W}_2^{[2]}\right) 
		\s 
		\left. + 8 \, \mbf F_4\left(\bs \phi_1^{[1]}, \bs \phi_1^{[1]}, \bs \phi_1^{[1]}, 3 \, \evs{\mbf W}_0^{[2]} + 2 \, \evs{\mbf W}_2^{[2]}\right) + 10 \, \mbf F_5\left(\bs \phi_1^{[1]}, \bs \phi_1^{[1]}, \bs \phi_1^{[1]}, \bs \phi_1^{[1]}, \bs \phi_1^{[1]}\right)\right). \label{eq:C5}
	\end{multline}

\subsection{Unfolding}
    Last, but not least, note that everything we have done so far is a linear process; we went order by order to find the coefficients of a polynomial, and was specific to the evaluation at $\varepsilon = 0$. To analyze what happens as $\varepsilon$ changes, we need to include an extra second-order term. In particular, we want $A$ to solve a differential equation with the following form:
	\begin{align}
		\partial_t A = h^{[1, 1]}(A, \varepsilon) + h^{[1, 0]}(A, \varepsilon) + h^{[2, 0]}(A, \varepsilon) + \ldots, \label{genAwithpar}
	\end{align}
    where the superscripts stand for the \evs{degree} of each polynomial in $A$ and $\varepsilon$, respectively. In particular, note that for each $i\in \mathbb Z^+$, $h^{[i, 0]}(A, \varepsilon) = h^{[i]}(A)$, which corresponds to one of the terms we calculated before. Now, for $A$ to solve the equation \eqref{genAwithpar}, we need to look for a change of variables of the form
	\begin{align*}
		\mbf u = \evs{\mbf u}^{[1, 1]}(A, \varepsilon) + \evs{\mbf u}^{[1, 0]}(A, \varepsilon) + \evs{\mbf u}^{[2, 0]}(A, \varepsilon) + \ldots
	\end{align*}
	Next, note that the terms $\evs{\mbf u}^{[i, 0]}(A, \varepsilon)$ for $i = 1, 2, 3, 4, 5$ were already found with our previous procedure. We are now interested in $h^{[1, 1]}(A, \varepsilon)$. To find it, we can consider just the terms in \eqref{geneq} that have a product between $A$ and $\varepsilon$ when we perform a Taylor expansion of our vector function $\mbf f$ with respect to $\mbf u$ and $\varepsilon$. In that case, we have the following expression:
	\begin{align*}
		\partial_A \evs{\mbf u}^{[1, 0]} \, h^{[1, 1]} + \mbox{conj.} = \left(\jac \mbf f(\mbf 0) + \mathbb D \, \partial_{xx}\right)\evs{\mbf u}^{[1, 1]} + \varepsilon \, \mbf F_{1, 1}\left(\evs{\mbf u}^{[1, 0]}\right) + \varepsilon \, \frac{\dd \mathbb D}{\dd \varepsilon} (0) \, \partial_{xx} \evs{\mbf u}^{[1, 0]},
	\end{align*}
	which is equivalent to
	\begin{align*}
		\left(\jac \mbf f(\mbf 0) + \mathbb D \, \partial_{xx}\right)\evs{\mbf u}^{[1, 1]} = e^{ikx} \, h^{[1, 1]} \, \bs \phi_1^{[1]} - A \, \varepsilon \, e^{ikx} \, \mbf F_{1, 1}\left(\bs \phi_1^{[1]}\right) + A \, \varepsilon \, e^{ikx} \, k^2 \, \frac{\dd \mathbb D}{\dd \varepsilon}(0) \, \bs \phi_1^{[1]} + c.c.
	\end{align*}
	Again, in this case\evs{,} we have a solvability condition that can be used to find $h^{[1, 1]}$. In particular, we can use the Fredholm Alternative with the vector function $\bs{\hat \psi}_+$. Note that
	\begin{align*}
		\innerp{\left(\jac \mbf f(\mbf 0) + \mathbb D \, \partial_{xx}\right)\evs{\mbf u}^{[1, 1]}}{\bs{\hat \psi}_+} = \innerp{e^{ikx} \, h^{[1,1]} \, \bs \phi_1^{[1]}}{e^{ikx} \, \bs \psi_1^{[0]}} - A \, \varepsilon \innerp{e^{ikx} \, \mbf F_{1,1}\left(\bs \phi_1^{[1]}\right)}{e^{ikx} \bs \psi_1^{[0]}}
		\s 
		+ A \, \varepsilon \, k^2 \innerp{e^{ikx} \frac{\dd \mathbb D}{\dd \varepsilon} (0) \, \bs \phi_1^{[1]}}{e^{\pm ikx} \, \bs \psi_1^{[0]}}
	\end{align*}
	Therefore, when we set this expression to zero, we can see that
	\begin{align*}
		h^{[1, 1]}(A, \varepsilon) = C_{1, 1} \, A \, \varepsilon,
	\end{align*}
	where
	\begin{align*}
		C_{1, 1} = \frac{1}{\bs \psi_1^{[0]}\cdot \bs \phi_1^{[1]}} \, \bs \psi_1^{[0]} \cdot \left( \mbf F_{1,1}\left(\bs \phi_1^{[1]}\right) - k^2 \, \frac{\dd \mathbb D}{\dd \varepsilon}(0) \, \bs \phi_1^{[1]} \right).
	\end{align*}
	Thus we have computed the final expression in the Theorem, and we have an amplitude equation of the form \eqref{eq:NFeqn}. 

    This concludes the proof of the main result.

\subsection{An extended normal form}\label{sec:3.6}
    The process so far computes the minimal normal form necessary for finding the criticality of a pattern formation, or Turing, bifurcation. Essentially, one is computing the normal form of the equivalent pitchfork bifurcation. Nevertheless, if one is interested in the complete dynamics of the amplitude in a neighbourhood of the bifurcation, then there are additional terms that arise at the same order in a weakly nonlinear expansion that leads to an extended form of Ginzburg-Landau equation; see for example \cite{ponedel2017front}. Such forms do not change the criticality of the bifurcation but contain additional terms such as a second spatial derivative of the amplitude, that give information on the dynamics of the amplitude equation up to order 5 when working near a codimension-two bifurcation point where $C_3\approx 0$. To complete this calculation in full, various assumptions need to be made about the order in which parameters change --- see for example \arc{\cite{Dean}}. Here, to remain general, we shall not compute the unfolding, and shall assume that we are working precisely at the codimension-two point where $C_3 = 0$. From now on, $\varepsilon$ will be considered as a parameter of the expansion.
        
    Let $0 < \varepsilon \ll 1$ be a small parameter. We define two extra spatial and temporal scales $X = \varepsilon^2 \, x$ and $T = \varepsilon^4 \, t$, respectively, that will be considered independent of $x$ and $t$. This implies that \eqref{geneq} becomes
    \begin{align}
        \partial_t \mbf u + \varepsilon^4 \, \partial_T \mbf u &= \mbf f(\mbf u,\varepsilon) + \mathbb D(\varepsilon) \, \left(\partial_{xx} \mbf u + 2 \, \varepsilon^2 \, \partial_{xX} \mbf u + \varepsilon^4 \, \partial_{XX} \mbf u\right). \label{eq:CGL}
    \end{align}
    Following the same idea we followed before, assuming that the solution $\mbf u$ changes only in the small time scale, \arc{w}e expand $\mbf u$ in the following way:
    \begin{align*}
        \mbf u = \sum_{i = 1}^N \varepsilon^i \, \mbf u^{[i]},
    \end{align*}
    which implies that
    \begin{align*}
        \mbf u^{[1]} = A(T, X) \, e^{ikx} \, \bs \phi_1^{[1]} + c.c.,
    \end{align*}
    and
    \begin{align*}
        \mbf u^{[2]} = \abs{A}^2 \, \mbf W_0^{[2]} + A^2 \, e^{2ikx} \, \mbf W_2^{[2]} + c.c.
    \end{align*}
    The difference between the previous approach and this one comes at order $\mathcal O\left(\varepsilon^3\right)$, where the expansion of \eqref{geneq} becomes
    \begin{align*}
        \mbf 0 = \left(\jac \mbf f(\mbf 0) + \mathbb D \, \partial_{xx}\right)\evs{\mbf u}^{[3]} + 2 \, \mbf F_2\left(\evs{\mbf u}^{[1]}, \evs{\mbf u}^{[2]}\right) + \mbf F_3\left(\evs{\mbf u}^{[1]}, \evs{\mbf u}^{[1]}, \evs{\mbf u}^{[1]}\right) + 2 \, \mathbb D \, \partial_{xX} \mbf u^{[1]}.
    \end{align*}
    Now, the solution is given by
    \begin{align*}
        \mbf u^{[3]} = \abs{A}^2 A \, e^{ikx} \, \mbf W_1^{[3]} + i \, \partial_X A \, e^{ikx} \, \mbf W_{1,2}^{[3]} + A^3 \, e^{3ikx} \, \mbf W_3^{[3]},
    \end{align*}
    \arc{where $\mbf W_1^{[3]}$ and $\mbf W_{1,2}^{[3]}$ solve} 
    \begin{align}
        \left(\jac \mbf f(\mbf 0) - k^2 \, \mathbb D\right) \mbf W_1^{[3]} = - 2 \, \mbf F_2\left(\bs \phi_1^{[1]}, 2 \, \mbf W_0^{[2]} + \mbf W_2^{[2]}\right) - 3 \, \mbf F_3\left(\bs \phi_1^{[1]}, \bs \phi_1^{[1]}, \bs \phi_1^{[1]}\right), \label{W13}
    \end{align}
    and
    \begin{align}
        \left(\jac \mbf f(\mbf 0) - k^2 \, \mathbb D\right) \mbf W_{1,2}^{[3]} = - 2 \, k \, \mathbb D \, \bs \phi_1^{[1]}\arc{,} \label{W123}
    \end{align}
    respectively, and $\mbf W_3^{[3]}$ is the same vector as before. \arc{Note that \eqref{W13} has a solution because we are at the codimension-two point, whereas \eqref{W123} is solvable because the critical wavenumber $k$, at a Turing bifurcation, is a double route of the dispersion curve $\Re(\lambda(k))$.}

    \evs{Next, at order $\mathcal O\big(\varepsilon^4\big)$, we have}
    \begin{multline}
		\mbf 0 = \left(\jac \mbf f(\mbf 0) + \mathbb D \, \partial_{xx}\right)\mbf u^{[4]} + 2 \, \mbf F_2\left(\mbf u^{[1]}, \mbf u^{[3]}\right) + \mbf F_2\left(\mbf u^{[2]}, \mbf u^{[2]}\right)
		\s 
		+ 3 \, \mbf F_3\left(\mbf u^{[1]}, \mbf u^{[1]}, \mbf u^{[2]}\right) + \mbf F_4\left(\mbf u^{[1]}, \mbf u^{[1]}, \mbf u^{[1]}, \mbf u^{[1]}\right) + 2 \, \mathbb D \, \partial_{xX} \mbf u^{[2]}. \label{genfourthorder}
	\end{multline}
    Therefore,
    \begin{align*}
        \mbf u^{[4]} = \abs{A}^4 \, \mbf W_0^{[4]} + i \, \bar A \, \partial_X A \, \mbf W_{0,2}^{[4]} + \abs{A}^2 \, A^2 \, e^{2ikx} \, \mbf W_2^{[4]} + i \, A \, \partial_X A \, e^{2ikx} \, \mbf W_{2, 2}^{[4]} + \ldots + c.c.,
    \end{align*}
    where
    \begin{multline*}
		\jac \mbf f(\mbf 0) \, \mbf W_0^{[4]} = - 2 \, \mbf F_2\left(\bs \phi_1^{[1]}, \mbf W_1^{[3]}\right) - 2 \, \mbf F_2\left(\mbf W_0^{[2]}, \mbf W_0^{[2]}\right) - \mbf F_2\left(\mbf W_2^{[2]}, \mbf W_2^{[2]}\right)
        \s
        - 3 \, \mbf F_3\left(\bs \phi_1^{[1]}, \bs \phi_1^{[1]}, 2 \, \mbf W_0^{[2]} + \mbf W_2^{[2]}\right) - 3 \, \mbf F_4\left(\bs \phi_1^{[1]}, \bs \phi_1^{[1]}, \bs \phi_1^{[1]}, \bs \phi_1^{[1]}\right),
	\end{multline*}
    \begin{align*}
		\jac \mbf f(\mbf 0) \, \mbf W_{0, 2}^{[4]} = - 2 \, \mbf F_2\left(\bs \phi_1^{[1]}, \mbf W_{1, 2}^{[3]}\right),
	\end{align*}
    \begin{multline*}
		\left(\jac \mbf f(\mbf 0) - 4 \, k^2 \, \mathbb D\right)\mbf W_2^{[4]} = - 2 \, \mbf F_2\left(\bs \phi_1^{[1]}, \mbf W_1^{[3]} + \mbf W_3^{[3]}\right) - 4 \, \mbf F_2\left(\mbf W_0^{[2]}, \mbf W_2^{[2]}\right)
		\s 
		- 6 \, \mbf F_3\left(\bs \phi_1^{[1]}, \bs \phi_1^{[1]}, \mbf W_0^{[2]} + \mbf W_2^{[2]}\right) - 4 \, \mbf F_4\left(\bs \phi_1^{[1]}, \bs \phi_1^{[1]}, \bs \phi_1^{[1]}, \bs \phi_1^{[1]}\right),
	\end{multline*}
    and
    \begin{align*}
		\left(\jac \mbf f(\mbf 0) - 4 \, k^2 \, \mathbb D\right)\mbf W_{2, 2}^{[4]} = - 2 \, \mbf F_2\left(\bs \phi_1^{[1]}, \mbf W_{1, 2}^{[3]}\right) - 8 \, k \, \mathbb D \, \mbf W_2^{[2]}.
	\end{align*}

    Finally, at order $\mathcal O\big(\varepsilon^5\big)$, \eqref{geneq} becomes
    \begin{multline}
		\partial_T \mbf u^{[1]} = \left(\jac \mbf f(\mbf 0)+\mathbb D \, \partial_{xx}\right) \mbf u^{[5]} + 2 \, \mbf F_2\left(\mbf u^{[1]}, \mbf u^{[4]}\right) + 2 \, \mbf F_2\left(\mbf u^{[2]}, \mbf u^{[3]}\right)
		\s 
		+ 3 \, \mbf F_3\left(\mbf u^{[1]}, \mbf u^{[1]}, \mbf u^{[3]}\right) + 3 \, \mbf F_3\left(\mbf u^{[1]}, \mbf u^{[2]}, \mbf u^{[2]}\right)
		\s 
		+ 4 \, \mbf F_4\left(\mbf u^{[1]}, \mbf u^{[1]}, \mbf u^{[1]}, \evs{\mbf u}^{[2]}\right) + \mbf F_5\left(\mbf u^{[1]}, \mbf u^{[1]}, \mbf u^{[1]}, \mbf u^{[1]}, \mbf u^{[1]}\right), \label{genfifthorder}
	\end{multline}
    which needs a solvability condition to ensure it has a solution. In particular, this implies that
    \begin{align*}
        \partial_T A = \alpha_1 \partial_{XX} A + i \, \alpha_2 \, \abs{A}^2 \, \partial_X A + i \, \alpha_3 \, A^2 \, \overline{\partial_X A} + \alpha_4 \, \abs{A}^4 \, A,
    \end{align*}
    where
    \begin{align*}
        \alpha_1 &= - \frac{2 \, k}{\bs \psi_1^{[0]} \cdot \bs \phi_1^{[1]}} \, \bs \psi_1^{[0]} \cdot \left(\mathbb D \, \mbf W_{1, 2}^{[3]}\right),
        \\
        \alpha_2 &= \frac{2}{\bs \psi_1^{[0]} \cdot \bs \phi_1^{[1]}} \, \bs \psi_1^{[0]} \cdot \left(\mbf F_2\left(\bs \phi_1^{[1]}, \mbf W_{0, 2}^{[4]} + \mbf W_{2, 2}^{[4]}\right) + 2 \, \mbf F_2\left(\mbf W_0^{[2]}, \mbf W_{1, 2}^{[3]}\right)\right.
        \\
        & \quad \left. + 3 \, \mbf F_3\left(\phi_1^{[1]}, \phi_1^{[1]}, \mbf W_{1, 2}^{[3]}\right) + 2 \, k \, \mathbb D \, \mbf W_1^{[3]}\right),
        \\
        \alpha_3 &= \frac{1}{\bs \psi_1^{[0]} \cdot \bs \phi_1^{[1]}} \, \bs \psi_1^{[0]} \cdot \left(- 2 \, \mbf F_2\left(\bs \phi_1^{[1]}, \mbf W_{0, 2}^{[4]}\right) - 2 \, \mbf F_2\left(\mbf W_2^{[2]}, \mbf W_{1, 2}^{[3]}\right)\right.
        \\
        & \quad \left. - 3 \, \mbf F_3\left(\bs \phi_1^{[1]}, \bs \phi_1^{[1]}, \mbf W_{1, 2}^{[3]}\right) + 2 \, k \, \mathbb D \, \mbf W_1^{[3]}\right)
        \\
        \alpha_4 &= \frac{1}{\bs \psi_1^{[0]} \cdot \bs \phi_1^{[1]}} \, \bs \psi_1^{[0]}\cdot \left(2 \, \mbf F_2\left(\bs \phi_1^{[1]}, 2 \, \mbf W_0^{[4]} + \mbf W_2^{[4]}\right) + 4 \, \mbf F_2\left(\mbf W_0^{[2]}, \mbf W_1^{[3]}\right)\right.
		\s 
		& \quad + 2 \, \mbf F_2\left(\mbf W_2^{[2]}, \mbf W_1^{[3]} + \mbf W_3^{[3]}\right) + 3 \, \mbf F_3\left(\bs \phi_1^{[1]}, \bs \phi_1^{[1]}, 3 \, \mbf W_1^{[3]} + \mbf W_3^{[3]}\right)
		\s 
		& \quad + 12 \, \mbf F_3\left(\bs \phi_1^{[1]}, \mbf W_0^{[2]}, \mbf W_0^{[2]} + \mbf W_2^{[2]}\right) + 6 \, \mbf F_3\left(\bs \phi_1^{[1]}, \mbf W_2^{[2]}, \mbf W_2^{[2]}\right)
		\s 
		& \quad \left. + 8 \, \mbf F_4\left(\bs \phi_1^{[1]}, \bs \phi_1^{[1]}, \bs \phi_1^{[1]}, 3 \, \mbf W_0^{[2]} + 2 \, \mbf W_2^{[2]}\right) + 10 \, \mbf F_5\left(\bs \phi_1^{[1]}, \bs \phi_1^{[1]}, \bs \phi_1^{[1]}, \bs \phi_1^{[1]}, \bs \phi_1^{[1]}\right)\right).
    \end{align*}
    Note that $\alpha_4$ is nothing other than the coefficient $C_5$ given by \eqref{eq:C5}. The parameters $\alpha_1$, $\alpha_2$, and $\alpha_3$ are also computed in the software implementation explained below. But as already stated, only $C_3$, given by \eqref{eq:C3} is required to compute the criticality of the bifurcation, and $C_5$ ($\equiv \alpha_4$) is required only if $C_3 \approx 0$.

\section{Computational details} \label{sec:comp}
    \evs{The main purpose of this paper is to provide users with an algorithm that automatically computes the coefficients of the Turing bifurcation normal form for a given example system. After setting up the definition of the various functions and parameters, the main computational effort is the computation of the coefficients $C_3$ and $C_5$. Algorithms \ref{algorithm1} and \ref{algorithm2} explain the main steps in pseudo-code. The main difference between these two algorithms is that one works using Python for the symbolic computations and Mathematica for plotting whilst the second one uses Mathematica to do the computation of the coefficients and plotting. We have tested both algorithms and they yield the same results in terms of the graphs and values of the fifth-order coefficient. The code in Mathematica uses all the symbolic power of Mathematica to do the calculations, but the code with Python can be more efficient because of Python's efficiency.}
    \subsection{Python + Mathematica}
    	\begin{algorithm}
    	    \caption{Determining $C_3$, $C_5$, \evs{$\alpha_1$, $\alpha_2$ and $\alpha_3$}}
            \label{algorithm1}
    	    \begin{algorithmic}
    	        \REQUIRE file ``functions.py'', folder ``model'' and file ``model.py'' inside of that folder
    	        \FOR{$x$ in var, parameters}
    	        \STATE $x \leftarrow \text{symbols}(x)$
    	        \ENDFOR
    	        \FOR{$x$ in kinetics, diffmatrix}
    	        \STATE $x \leftarrow \text{eval}(x)$
    	        \ENDFOR
    	        \STATE jacobianmat$\leftarrow$ diff(kinetics,var)$-\mu \cdot $ diffmatrix
    	        \STATE jacobianmatdet$\leftarrow$det(jacobianmat)
    	        \STATE determinantderivative $\leftarrow$ diff(jacobianmatdet,$\mu$)
    	        \STATE $\mu \leftarrow$ solve(\{jacobianmatdet,determinantderivative\},$\mu$)
    	        \STATE $\phi_1^{[1]} \leftarrow$ solve(jacobianmat $\cdot \phi_1^{[1]}=\mbf 0$ ,$\phi_1^{[1]}$)
    	        \STATE solve linear systems to determine second-order vectors
    	        \STATE $\psi_1^{[0]} \leftarrow$ solve(jacobianmat$^\intercal \cdot \psi_1^{[0]}=\mbf 0$ ,$\psi_1^{[0]}$)
    	        \STATE compute $C_3$
    	        \STATE solve linear equations to find third and fourth-order vectors
    	        \STATE compute $C_5$, \evs{$\alpha_1$, $\alpha_2$ and $\alpha_3$}
    	        \STATE save text files with the data required 
    	    \end{algorithmic}
    	\end{algorithm}
    	
        We have implemented Algorithm \ref{algorithm1} as a Python script that is available \evs{in} \url{https://github.com/edgardeitor/TOMS}. There, \evs{we also put} a Mathematica script that can take the output of the Python code and is able to graph the bifurcation curves and their criticality and to automatically detect codimension-two points at which $C_{\evs{3}} = 0$. The code has been tested using Mathematica version \evs{13.2} and Python version \evs{3.10.9}.
        
        To run the Python script, it is necessary to have six packages in Python: {\tt IPython}, {\tt os}, {\tt shutil}, {\tt sys}, {\tt sympy} and {\tt mpmath}.
    	
        We recommend creating a core directory to place the files {\tt Criticality.py}, {\tt functions.py} and {\tt Plotter.nb}. Besides, create a sub-directory for each model to analyze, with the name {\tt foo} of the system and a file {\tt foo.py} using the same name as that of the sub-directory. We present a brief description of these four files.
    	\begin{description}
    	    \item[{\tt Criticality.py}] This file has the main part of the code, which computes all the derivatives and obtains the formulas for each of the vectors needed to determine the third and fifth-order coefficients of the normal form.
    	    \item[{\tt foo.py}] This file has all the information about the reaction-diffusion equation. One must write all the required information asked in that file if one wants to run the main script properly. The description of each required variable can be found in each file.
    	    \item[{\tt functions.py}] This file contains all the functions and classes that the main script needs to compute \evs{the} criticality of the bifurcation.
    	    \item[{\tt Plotter.nb}] This file contains the code required to obtain the bifurcation curves numerically and plot them using colors to identify the criticality of the bifurcation at each point.
    	\end{description}
            
        To run the code, \evs{one} simply needs to edit {\tt foo.py} and then run the main Python script. If one provides all the information correctly, this code will need just one mandatory input, and more input in a few cases: first, one needs to provide the name of the folder \arc{with the model}. Apart from that, if there is more than one real homogeneous steady-state in \evs{the} system, the script will ask to choose the steady state \evs{one} wants to study. If Python does not find a real homogeneous steady-state for \evs{the} system, the script will ask to provide a parameter that can be changed in order to find a real homogeneous steady-state. \evs{One can also tell Python that \arc{one does} not want to change any parameters and the symbolic computation will still continue, but there is no guarantee that Mathematica will find any Turing bifurcation after}.
    	
        After running the main Python script, new files \evs{will} appear in the folder; some new text files and, if the appropriate parameters were specified, the file {\tt Plotter.nb}. If \evs{one} wants to graph the Turing bifurcation curves, \evs{one} then needs to open {\tt Plotter.nb} with Mathematica and run all the cells in order. This can be done manually or \evs{one} can just set it to run the whole Notebook. Mathematica will ask for input only in the case that the system has more than one homogeneous-steady state, in which case it will ask to enter the \evs{numerical label} of the equilibrium \evs{one wants} to analyze. This also has the option to input a number 0 if \evs{one wants} to analyze all the equilibria of the system at the same time. As Mathematica is a stronger software with symbolic expressions than Python, these equilibria will be provided symbolically. For that reason, the \evs{numerical label} of the homogeneous steady-state \evs{one wants} to study needs to be provided again (it \evs{may} be different from the \evs{label} of the \evs{equilibrium} in Python). Also, if \evs{one wants}, \evs{one} can change the equilibrium considered in Python to see what can be found for each equilibrium in the same rectangle \evs{one is} using to plot the bifurcation curves.
    	
        The curves are plotted using colors according to what they are. In particular, red (respectively, blue) lines represent supercritical (respectively, subcritical) Turing bifurcation curves, while green dotted curves are \evs{the} so-called Belyakov-Devaney (BD) lines, in which the Turing bifurcation conditions are met, but $\mu = k^2<0$. Besides, gray curves are fold curves where $\det(\jac \mbf f(\mbf P)) = 0$.
        
        We highlight here that BD lines are not bifurcation curves as such (for that reason they are shown \evs{as} a dotted line), but it is important to locate them because they are curves in which the eigenvalues of the spatial reaction-diffusion equation change from real to complex, marking a change between oscillatory and non-oscillatory behavior of solutions in the boundary of the domain $[-L, L]$ \cite{dissecting}.
        
        Details of the numerical methods used in {\tt Criticality.py} and {\tt Plotter.nb} are given in Appendix \ref{appendixA}.
        
        The code has been tested for at least twelve different models, and the default parameter values given in the scripts work well for most of them. If a user finds any problems with the parameters or the execution of the code for a particular system, suggestions can be made through the GitHub repository \url{https://github.com/edgardeitor/TOMS}.
    \subsection{Mathematica only}
        \evs{We have implemented Algorithm \ref{algorithm2} as a Mathematica script that is available in \url{https://github.com/edgardeitor/TOMS}. For this algorithm, there is only one file needed, which is called ``Plotter.nb''. The only thing necessary for the script to run is to set the parameters in the second section and the kinetics and diffusion matrix in the third one. Though this seems to be simpler, this code may take more time to run as it performs the Taylor expansion of the vector field symbolically. The results of both scripts are the same and the examples are provided with both versions of the code.}
	  \begin{algorithm}
    	    \caption{Determining $C_3$ and $C_5$}
            \label{algorithm2}
    	    \begin{algorithmic}
    	        \REQUIRE file ``plotter.nb''
                \STATE jacobianmat$\leftarrow$ diff(kinetics, var)$-\mu \cdot $ diffmatrix
    	        \STATE jacobianmatdet$\leftarrow$det(jacobianmat)
    	        \STATE determinantderivative $\leftarrow$ diff(jacobianmatdet,$\mu$)
                \STATE $\mu \leftarrow$ solve(\{jacobianmatdet,determinantderivative\},$\mu$)
    	        \STATE $\phi_1^{[1]} \leftarrow$ solve(jacobianmat $\cdot \phi_1^{[1]}=\mbf 0$ ,$\phi_1^{[1]}$)
    	        \STATE $\psi_1^{[0]} \leftarrow$ solve(jacobianmat$^\intercal \cdot \psi_1^{[0]}=\mbf 0$ ,$\psi_1^{[0]}$)
                \STATE Obtain bifurcation curves
    	        \FOR{$x$ in var}
    	        \STATE $x \leftarrow \sum_{i=0}^{N} \varepsilon^i x_i$
    	        \ENDFOR
    	        \FOR{$i$ in \{1,\ldots, N\}}
    	        \STATE solve linear equations to find gotten through the expansion
    	        \ENDFOR
    	        \STATE compute $C_3$ and $C_5$ (if requested)
    	        \STATE save data files and plot bifurcation curves colored according to the sign of $C_3$ and $\mu$
    	    \end{algorithmic}
    	\end{algorithm}
	
\section{Examples} \label{sec:examples}
    Scripts to run the examples in each of the following subsections are provided with the code.
      	
    \subsection{Schnakenberg model with cross-diffusion.}
\begin{figure}[tbp]
		\centering
		\includegraphics[scale=0.5]{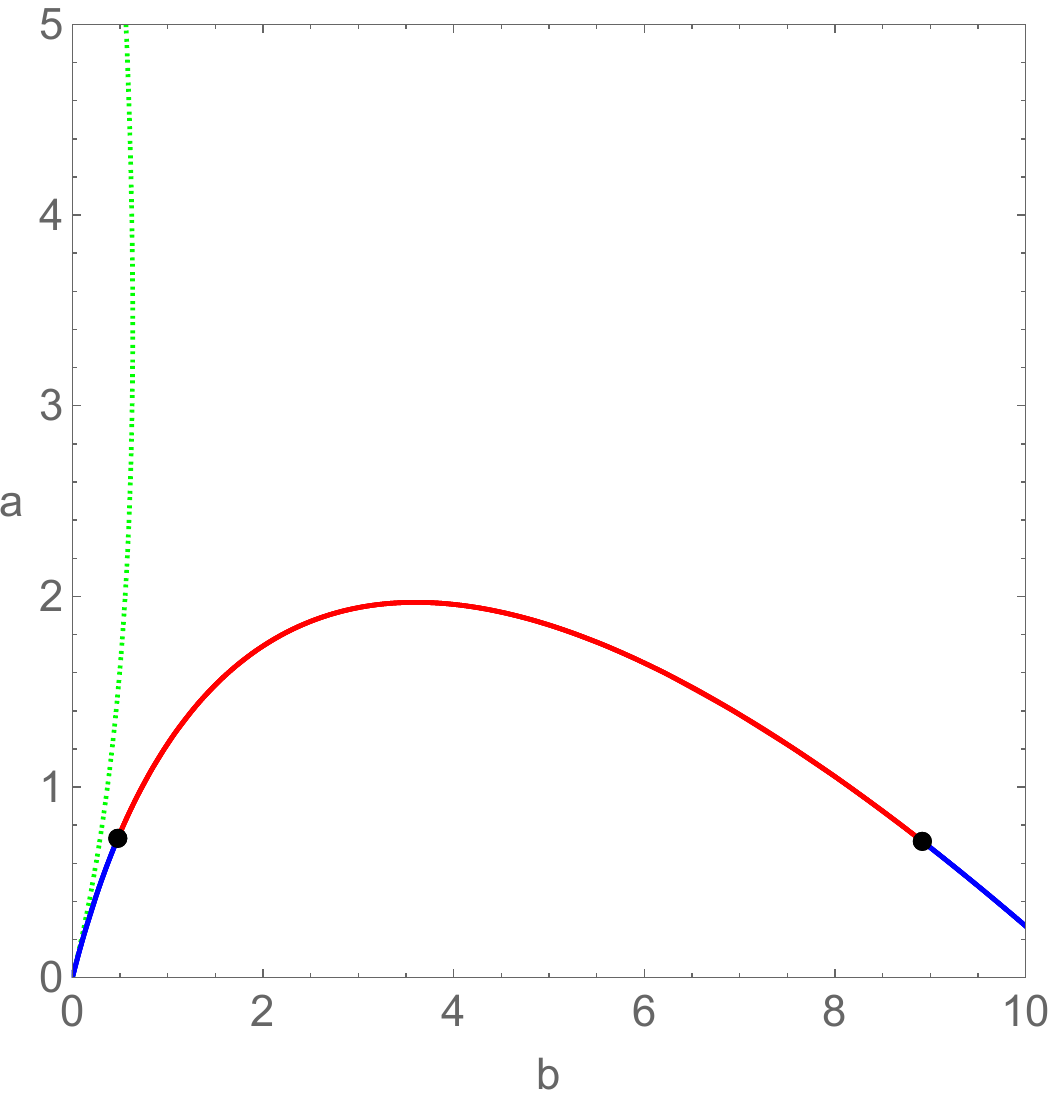}
		\caption{Bifurcation diagram of the Schnakenberg model with the addition of cross-diffusion. The dotted green curve represents a BD-line and the blue (resp. red) curve is a subcritical (resp. supercritical) Turing bifurcation. The black points are codimension-two bifurcation points where the third-order coefficient becomes zero.}
		\label{fig:schnakenberg}
	\end{figure}

    To extend previous studies on the Schnakenberg model (see e.g.~\cite{fahad}) we add extra terms that take into account the phenomenon of cross-diffusion between the variables of the system in the following way:
	\begin{align*}
		u_t&=a-u+u^2v+\delta^2 u_{xx}+h_1 v_{xx},
		\s 
		v_t&=b-u^2v+h_2 u_{xx} + v_{xx}.
	\end{align*}
	This model has only one homogeneous steady state
	\begin{align*}
		\mbf P&=\left(a+b,\frac{b}{(a+b)^2}\right).
	\end{align*}
	After doing all the computations of the terms in order to characterize the bifurcation and evaluating the following diffusion parameters
	\begin{align*}
		\left(\delta, h_1, h_2\right) = \left(0.07, 0.5, -0.001\right),
	\end{align*}
    we obtain the bifurcation diagram shown in figure \ref{fig:schnakenberg}. There, we can see two codimension-two bifurcation points where the third-order coefficient of the Turing bifurcation becomes zero. In both cases, the fifth-order coefficient is negative. This lets us characterize the Turing bifurcation completely at each of the points of the diagram.
	
	The files to run this example and obtain the graph shown are in GitHub, by the name of ``Schnakenberg model with cross-diffusion''.

	\subsection{Swift-Hohenberg Model.}
        The Swift-Hohenberg model is a fourth-order equation with only one component given by
	\begin{align}
	    \partial_t u = r u - \left(\partial_x^2 + q_c^2\right)^2 u + f(u). \label{Swift-Hohenberg}
	\end{align}
    See \cite{Knobloch_review} and references therein for what is known about localised pattern-formation in this equation with bistable kinetic terms $f(u)$. Though this equation is not a reaction-diffusion system of the form \eqref{geneq}, we explain how the method can nevertheless be used.

    In general, consider a scalar reaction-diffusion equation of the form 
	\begin{align}
	  \partial_t u&=g\left(u,\partial_x^2 u, \partial_x^4 u,\ldots ,\partial_x^{2m}u\right), \label{manyspatialderivatives}
        \end{align}
        where 
	\begin{align*}
		g\left(u,v_1,v_2,\ldots ,v_{m-1},\partial_x^2 v_{m-1}\right)=h\left(u,v_1,v_2,\ldots ,v_{m-1}\right)+L\left(\partial_x^2 v_{m-1}\right),
	\end{align*}
    with $L$ a linear function. Then can define a differential algebraic system by means of the auxiliary variables $v_1,v_2,\ldots ,v_{m-1}$ as follows: 
	\begin{align*}
		\partial_t u&=g\left(u,v_1,v_2,\ldots ,v_{m-1},\partial_x^2 v_{m-1}\right),
		\s 
		0&=v_1-\partial_x^2 u,
		\s 
		0&=v_2-\partial_x^2 v_1,
		\s 
		&\vdots
		\s 
		0&=v_{m-1}-\partial_x^2 v_{m-2}.
	\end{align*}
	With this, the formulation we developed can be used in the same way. The only thing we need to take extra care of is that the last $m-1$ equations need to have a zero derivative with respect to $t$, so we need to enforce the last $m-1$ components of the left-hand side of equations such that \eqref{thirdordereq} to be equal to 0.
	
    In particular, the Swift-Hohenberg equation \eqref{Swift-Hohenberg} can be written as
	\begin{align*}
	    \partial_t u &= r u - q_c^4 u - 2 q_c^2 v - v_{xx} + f(u),
	    \\
	    0 &= v - u_{xx}.
	\end{align*}

    We illustrate the computation for the case
    $$
        f(u) = b_3 u^3 - b_5 u^5,
    $$
    choosing the homogeneous steady-state given by $\mbf P = (u^*,v^*) = (0,0)$. This steady state goes through a Turing bifurcation when $r=0$ for $\arc{k} = q_c$. In particular, when running the script with this model, we get
	\begin{align*}
	    C_3 &= 3 b_{3} - 30 b_{5} u^{2} - \frac{\left(6 b_{3} u - 20 b_{5} u^{3}\right)^{2}}{2 \left(3 b_{3} u^{2} - 5 b_{5} u^{4} - 4 \arc{k^2} \left(4 \arc{k^2} - 2 q_{c}^{2}\right) - q_{c}^{4} + r\right)} - \frac{\left(6 b_{3} u - 20 b_{5} u^{3}\right)^{2}}{3 b_{3} u^{2} - 5 b_{5} u^{4} - q_{c}^{4} + r}.
	\end{align*}
    The fifth-order coefficient is a cumbersome expression that we omit here for the sake of brevity. Evaluation of these coefficients at $r=0$, we get the following expressions
	\begin{align*}
	    C_3 &= 3b_3, \qquad C_5 = \frac{3 b_3^2}{64 q_c^4} - 10 b_5,
	\end{align*}
    which are the same expressions as obtained in \cite[Eq.(28)]{knobloch}.

	\subsection{Konishi and Hara's model.} 
\begin{figure}[tbp]
		\centering
		\includegraphics[scale=0.5]{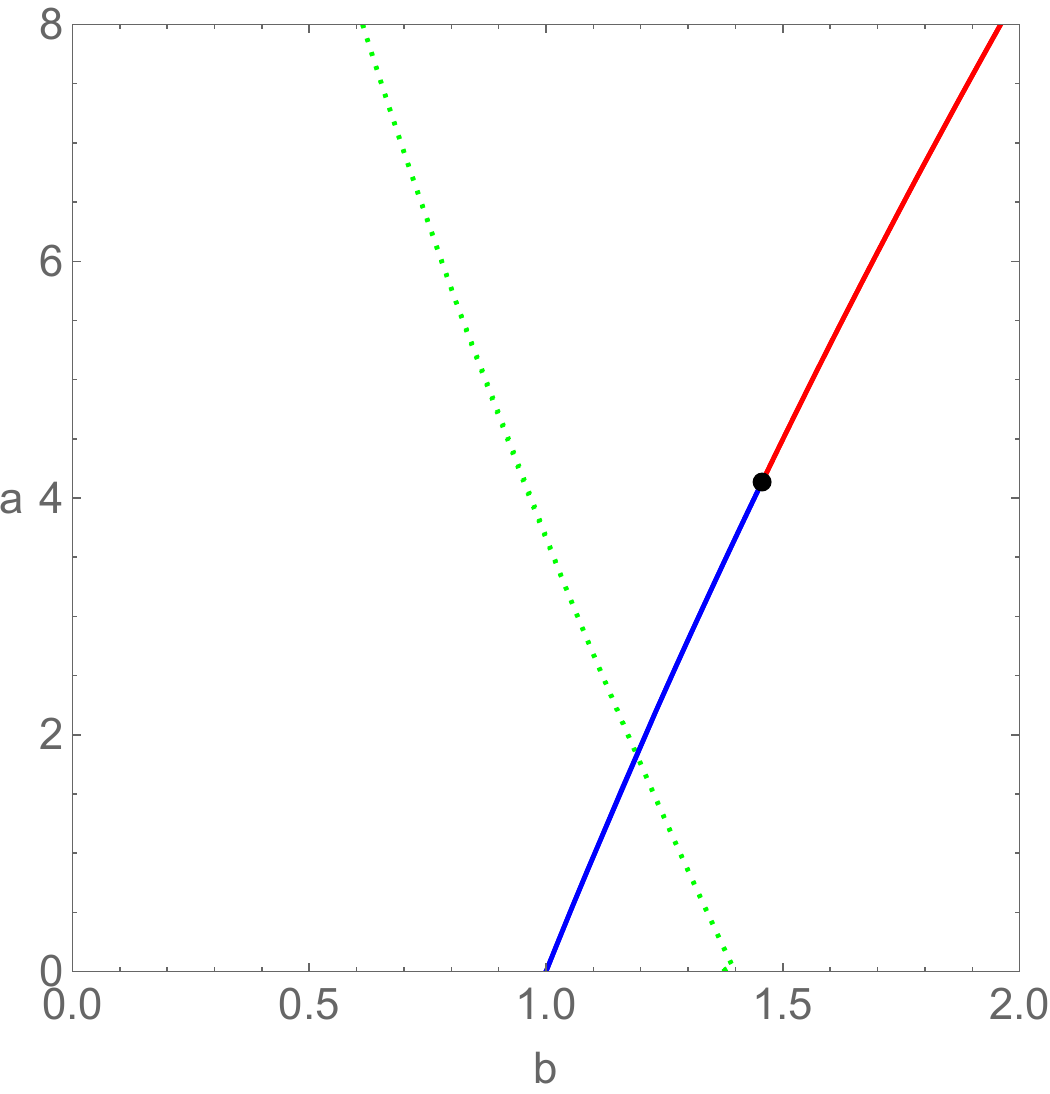}
		\caption{Bifurcation diagram of Konishi and Hara's model. The green dotted curve represents a BD-line and the blue (resp. red) curve is a subcritical (resp. supercritical) Turing bifurcation. The black \evs{point is a codimension-two bifurcation point} where the third-order coefficient becomes zero.}
		\label{fig:konishitb}
	\end{figure}

Finally, we consider \evs{the} four-component model studied in \cite{konishi}:
	\begin{align*}
		u_t &= \eta \left(a - (b + 1) u + u^2 v\right) + c (w - u) + \delta u_{xx},
		\s 
		v_t &= \eta \left(- u^2 v + b u\right) + c (z - v) + v_{xx},
		\s 
		w_t &= \eta \left(a - (b + 1) w + w^2 z\right) + c (u - w) + \mu \delta w_{xx},
		\s 
		z_t &= \eta \left(- w^2 z + b w\right) + c (v - z) + \mu z_{xx}.
	\end{align*}
	This model has many equilibrium points but we are interested in the simplest one, which is given by
	\begin{align*}
		\mbf P &= \left(a, \, \frac{b}{a}, \, a, \, \frac{b}{a}\right).
	\end{align*}
	In particular, we take into account the following parameter values:
	\begin{align*}
		\left(c, \delta, \eta, \mu \right)&=\left(0.01, 0.0025, 0.05, 1\right),
	\end{align*}
    some of the Turing bifurcation curves we can find are shown in figure \ref{fig:konishitb}. Again, in this diagram, we have a codimension-two bifurcation point in which the fifth-order coefficient is negative.

    \subsection{Continuation of codimension-two bifurcations}
        
\begin{figure}[tbp]
		\centering
		\includegraphics[scale=0.5]{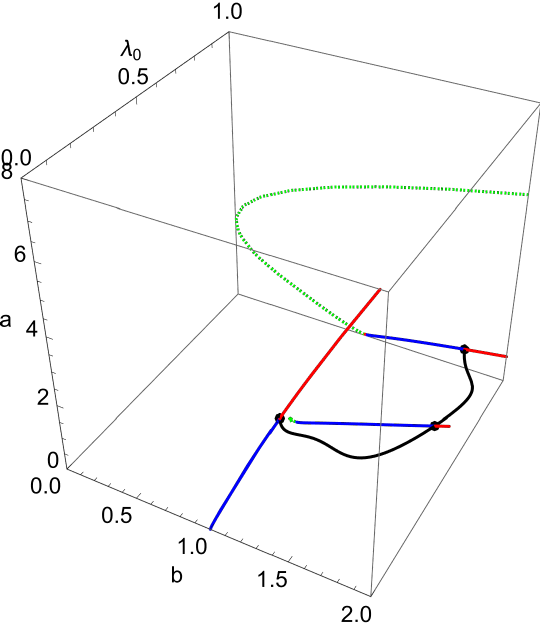}
		\caption{Codimension-2 bifurcation curve, together with some Turing bifurcation curves plotted along the way, together with the corresponding colors. The green dotted curve represents a BD-line and the blue (resp. red) curves are subcritical (resp. supercritical) Turing bifurcations. The black points are codimension-two bifurcation points where the third-order coefficient becomes zero.}
		\label{fig:konishicod2}
	\end{figure}

    Finally, we illustrate how to follow curves of the codimension-two bifurcation point as three parameters vary, by considering a modification of the previous example. In particular, let us define
	\begin{align*}
		f_1(x)&=\eta \left(a-(b+1)u +u^2v \right)+c (w-u) + \delta u_{xx},
		\s 
		f_2(x)&=\eta \left(-u^2v+bu\right)+c(z-v)+v_{xx},
		\s 
		f_3(x)&=\eta\left(a-(b+1)w+w^2z\right)+c(u-w)+\mu \delta w_{xx},
		\s 
		f_4(x)&=\eta\left(-w^2z+bw\right)+c(v-z)+\mu z_{xx},
		\s 
		g_1(x)&=\eta \left(a-(b+1)u +u^2v \right)+c_2 (w-u) + \delta_2\evs{^2} u_{xx},
		\s 
		g_2(x)&=\eta \left(-u^2v+bu\right)+c_2(z-v)+v_{xx},
		\s 
		g_3(x)&=\eta\left(a-(b+1)w+w^2z\right)+c_2(u-w)+\mu_2 \delta_2\evs{^2} w_{xx},
		\s 
		g_4(x)&=\eta\left(-w^2z+bw\right)+c_2(v-z)+\mu_2 z_{xx},
	\end{align*}
    with parameter values
    $$		
        c_2 =0.025, \quad 
		\delta_2 =0.2, \quad 
		\mu_2 =0.19,
    $$
    which were the values considered in \cite{konishi}. With this, we introduce a homotopy parameter $\lambda_0 \in [0,1]$ and consider the system 
	\begin{align*}
		u_t&=\lambda_0 f_1(x) + (1-\lambda_0) g_1(x),
		\s 
		v_t&=\lambda_0 f_2(x) + (1-\lambda_0) g_2(x),
		\s 
		w_t&=\lambda_0 f_3(x) + (1-\lambda_0) g_3(x),
		\s 
		z_t&=\lambda_0 f_4(x) + (1-\lambda_0) g_4(x).
	\end{align*}
    A plot of the locus of codimension-two bifurcation points is shown in figure \ref{fig:konishicod2}.

\begin{acks}
    E. V-S. has received PhD funding from ANID, Beca Chile Doctorado en el extranjero, number 72210071.
\end{acks}

\bibliographystyle{ACM-Reference-Format}
\bibliography{toms}


\newpage

\section*{Appendix}

\appendix

\section{Methods to compute and graph Turing bifurcation curves} \label{appendixA}

\subsection{``Criticality.py''}

    The Python script starts by running the file {\tt functions.py} and uses {\tt example.py} within the {\tt example} folder to obtain the number of variables of the system executing the code then defines each variable and parameter as a {\tt sympy} symbol and then it evaluates the {\tt Kinetics} at the parameter values provided, to find real homogeneous steady-states. After this, it computes the necessary functions that determine the conditions to have a Turing bifurcation curve, namely,
    \begin{align}
        f(\mu)=\det\left(\jac \mbf f(\mbf P)-\mu \, \mathbb D\right), \qquad \text{ and } \qquad g(\mu)=\frac{\partial f}{\partial \mu} (\mu),
        \label{eq:both}
    \end{align}
    where $\mu=k^2$. With these functions, it finds values of $\mu$ such that $g(\mu)=0$, then substitutes those values into $f$,  and checks whether any of those solutions solves the equation $f(\mu)=0$ within a tolerance of  {\tt tol}. After finding such a solution, the value of $\mu$ is saved and the code continues. Otherwise, the software asks the user to provide the name of a parameter of the system that can be changed to find a Turing bifurcation. Then, \evs{it} tries to find a solution to equations \eqref{eq:both} numerically using the function {\tt findroot} from {\tt mpmath}; solving both equations separately for each variable, or finding the resultant between $f$ and $g$. If none of these methods work, then it asks the user to provide a value of $k=\sqrt{\mu}$ manually.

    Then, Python computes the symbolic derivatives of the {\tt Kinetics} up to order five and saves them into arrays which then are used to optimize the code by avoiding Python from computing the same derivatives many times to compute the values of the functions $\mbf F_i$, for $i=2,3,4,5$. There is one last key ingredient to the computation. Note that the vector $\bs \phi_1^{[1]}$, specified through the equation
    \begin{align*}
        \left(\jac \mbf f(\mbf P)-\mu \, \mathbb D\right) \bs \phi_1^{[1]} = \mbf 0,
    \end{align*}
    is not uniquely defined and. Also, within Python, $\jac \mbf f(\mbf P)-\mu \, \mathbb D$ is a symbolic matrix. To circumvent this problem, let $c_1,c_2,\ldots ,c_n$ be column vectors of size $n$ such that
    \begin{align*}
    	\jac \mbf f(\mbf P)-k^2 \, \mathbb D=\begin{pmatrix}
    		\begin{array}{c|c|c|c}
    			c_1 & \cdots & c_{n-1} & c_n
    		\end{array}
    	\end{pmatrix} \qquad \bs \phi_1^{[1]}=\left(\phi_1,\phi_2,\ldots ,\phi_n\right)^\intercal.
    \end{align*}
    Then, note that
    \begin{align*}
    	\left(\jac \mbf f(\mbf P)-k^2 \, \mathbb D\right) \bs \phi_1^{[1]}&= \sum_{p=1}^n \phi_p \, c_p
    	\s 
    	&=\sum_{p=1}^{j-1} \phi_p \, c_p + \phi_j \, c_j + \sum_{p=j+1}^n \phi_p \, c_p.
    \end{align*}
    Therefore, given that $\jac \mbf f(\mbf P)-k^2 \, \mathbb D$ does not have full rank, there must exist positive integers $i$ and $j$ \evs{such that} we can remove the $i$-th row and $j$-th column of the matrix without changing the solution to $\left(\jac \mbf f(\mbf P)-k^2 \, \mathbb D\right) \bs \phi_1^{[1]}=\mbf 0$. The reduced equation is equivalent to
    \begin{align*}
    	B \, \bs{\hat \phi}_1^{[1]} +\phi_j \, \hat{c}_j&=\mbf 0,
    \end{align*}
    where
    \begin{align*}
    	B=\begin{pmatrix}
    		\begin{array}{c|c|c|c|c|c}
    			\hat c_1 & \cdots & \hat c_{j-1} & \hat c_{j+1} & \cdots & \hat c_n
    		\end{array}
    	\end{pmatrix} \qquad \bs{\hat \phi}=\left(\phi_1,\ldots ,\phi_{j-1},\phi_{j+1},\ldots , \phi_n\right)^\intercal,
    \end{align*}
    and $\hat c_k$ is the vector $c_k$ after \evs{removing} its $i$-th element. In particular, if we choose $\phi_j=1$, then the system is equivalent to
    \begin{align*}
    	B \, \bs{\hat \phi}_1^{[1]} + \hat c_j = \mbf 0.
    \end{align*}
    The solution of this system is \evs{now well-defined} provided $\dim\left(\ker\left(\jac \mbf f(\mbf P)-k^2 \, \mathbb D\right)\right)=1$.

    Thus, the code takes the matrix $\jac \mbf f(\mbf P)-k^2 \, \mathbb D$ and starts a search over $i$ and $j$ until it finds an invertible $(n-1)\times (n-1)$ submatrix.
    After this, each vector within the algorithm can be solved for uniquely, using the corresponding linear system of equations, using the above strategy only for finding $\bs \phi_1^{[1]}$, $\bs \psi_1^{[0]}$, and $\mbf Q_1^{[3]}$.

    Finally, Python just saves the important quantities into text files to leave everything ready for Mathematica to work.

\subsection{``Plotter.nb''} \label{Mathematica}
    The Mathematica script starts by defining auxiliary functions that will be used to format \LaTeX strings from the text files saved by Python. Then, it loads the file with the {\tt Kinetics} and computes a list of homogeneous steady states from it. If there is $p>1$ homogeneous steady state, the user will be asked to enter the index (from 1 to $p$) of the equilibrium that is required to be analyzed. A choice can also be made to consider all the homogeneous steady-states found.

    After that, Mathematica loads the bifurcation functions $f$ and $g$ and evaluates the fixed parameters at the values provided in the file {\tt exmample.py}. Mathematica then obtains bifurcation curves by following a numerical path-following approach. First, there is a parameter called {\tt simpleeq}, which can be set equal to 1 if the homogeneous steady-state under analysis is available in closed form, or 0 otherwise. Even if the equilibrium does not have a simple closed-form expression, the user can define {\tt simpleeq}$=1$ and Mathematica will attempt to find the equilibrium symbolically, but the computation could take longer or it may not finish. 

    Next, \evs{Mathematica} finds conditions under which a Turing bifurcation occurs. To do this, it fixes a parameter value (which is provided in the file {\tt name\_of\_folder.py}), and then, if {\tt simpleeq}$=0$ or no single equilibrium was chosen, it solves the equations
    $$
        f(\mu,p_1,p_2,\mbf u)=g(\mu,p_1,p_2,\mbf u)=\mbox{{\tt Kinetics}}(p_1,p_2,\mbf u)=\mbf 0
    $$
    numerically, where $p_1$ and $p_2$ are the parameters on the axes of the bifurcation diagram, with one of them being fixed. Otherwise, it solves the equations $f(\mu,p_1,p_2)=g(\mu,p_1,p_2)=0$ numerically, where one of the parameters is fixed and the chosen equilibrium is evaluated. Then, each solution to these equations is considered as an initial condition for the constrained ordinary differential equation (ODE)
    \begin{align*}
        \frac{\dd}{\dd s} \, \mbox{\tt Kinetics}(p_1(s),p_2(s),\mbf u(s))=\mbf 0, \quad \frac{\dd}{\dd s} f(\mu(s),p_1(s),p_2(s))=\frac{\dd}{\dd s} g(\mu(s),p_1(s),p_2(s))&=0,
        \\
        \norm{\mbf u'(s)}^2+p_1'(s)^2+p_2'(s)^2+\mu'(s)^2&=1.
    \end{align*}
    Here, the time constant of the evolution along the curve can be changed easily using the parameter {\tt curvespeed}.

    Next, to find codimension-two bifurcation points, Mathematica evaluates the third-order coefficient at each value of $s$ along the curve, identifying the changes of sign between one point and another. The point where there is a change of sign is then used as an initial condition to find the solution to the equation $C_3=0$ using the function {\tt FindRoot}, saving the coordinates of these points. Then, Mathematica plots the curves found -- colored according to the sign of $C_3$ -- and places black dots at the codimension-two bifurcation points.

    Curves of the codimension-two point in three parameters can be carried out using a similar procedure, by adding an extra equation
    $$
        \frac{\dd}{\dd s} C_3(\mu(s),p_1(s),p_2(s),\mbf u(s))=0
    $$
    to the ODE for finding the bifurcation curves. This completes the process. All relevant output is displayed in separate windows and the plotter, just in case \evs{one wants} to see the graphs without needing to run the plotter again.

    There are some additional parameters that the user can alter within the Mathematica script. These parameters can be found at the beginning of the cells called {\tt Vector field}, {\tt Bifurcation curves}, {\tt Finding codimension-two bifurcation points and computing \evs{the} fifth-order coefficient}, {\tt Codimension-two bifurcation curves} and {\tt Other Turing bifurcation curves}. These parameters are explained below:
    \begin{description}
        \item[{\tt eqbool}] This parameter must be equal to 1 if \evs{one wants} Mathematica to find algebraic expressions of the homogeneous steady-states and 0 otherwise.
        \item[{\tt initialt}, {\tt finalt}] These parameters are to define the domain of the functions $p_1(s),p_2(s),\mu(s),\mbf u(s)$, where $s\in I\subset [\text{\tt initialt},\text{\tt finalt}]$.
        \item[{\tt curvespeed}] As already explained, the user can set the value of the square root of $\norm{\mbf u'(t)}^2+p_1'(t)^2+p_2'(t)^2+\mu'(t)^2$ with this parameter.
        \item[{\tt tol}]  This defines the tolerance for numerical solution of an algebraic equation. It can take different values within different cells of the code. 
        \item[{\tt Choptol}]  This parameter is another tolerance that works together with the function {\tt Chop} that replaces sufficiently small numbers by an actual 0. Here, the parameter {\tt Choptol} defines the tolerance that indicates what `sufficiently small' is.
        \item[{\tt simpleeq}] This parameter defines whether the algebraic expression is simple enough to be evaluated directly within $f$ and $g$ to compute the initial conditions for a Turing bifurcation curve, or not.
        \item[{\tt smalldifferential}] \evs{The task} of the script is to graph bifurcation curves on a rectangle $(p_1,p_2)\in [a,b]\times [c,d]$. This parameter makes the computation \evs{stop when} $p_1$ (resp. $p_2$) is out of the interval $[a,b]$ (resp. $[c,d]$) by the distance {\tt smalldifferential}.
        \item[{\tt timedifference}] This variable defines the step-size in the `time' $s$ along a bifurcation curve, for the evaluation of $C_3$ used to find codimension-two bifurcation points.
        \item[{\tt disttol}] Because Mathematica uses many numerical tools, it is important for the code to recognise two sets of initial conditions for curves that are sufficiently close to be in reality the same point. This parameter sets the tolerance in distance to define two points as being equal.
    \end{description}

    In general, the user typically does not need to worry about the values of these parameters. They need only be changed if needed. Testing of the software has found that the most significant parameters that may need adjustment for a new example are {\tt tol} and {\tt simpleeq}. Also, to make the final output look sufficiently continuous and smooth, it may be necessary to adjust {\tt curvespeed}.

\end{document}